\documentclass[12pt,a4paper]{article}
\usepackage[margin=1in]{geometry}
\usepackage{setspace}
\usepackage[utf8]{inputenc}
\usepackage[T1]{fontenc}
\usepackage{microtype}
\usepackage{graphicx}
\usepackage{epstopdf}
\usepackage{xcolor}

\usepackage{amsmath,amssymb}
\usepackage{bm}
\usepackage{booktabs}
\usepackage{multirow}
\usepackage{caption}
\usepackage{subcaption}
\usepackage{siunitx}
\DeclareSIUnit\angstrom{\text{\AA}} 
\usepackage[version=4]{mhchem} 
\usepackage[colorlinks,urlcolor=black,citecolor=black,linkcolor=black]{hyperref}
\usepackage{cleveref}
\usepackage{authblk}
\usepackage[numbers,sort&compress]{natbib}

\title{Equivariant generative diffusion learns and generalizes the structural ensemble of amorphous oxides}

\author{Jun Jiang, Ian Berry, James N. Fry, and Hai-Ping Cheng\\[2pt]{\normalsize Correspondence: \href{mailto:j.jiang@northeastern.edu}{j.jiang@northeastern.edu}, \href{mailto:ha.cheng@northeastern.edu}{ha.cheng@northeastern.edu}}}
\affil{Department of Physics, Northeastern University, Boston, MA 02115, USA}
\date{\today}

\begin{document}
\maketitle

\begin{abstract}
Amorphous materials are statistical ensembles rather than definitive structures, and conventional density-functional (DFT) and machine-learned-potential simulations sample only a small part of that ensemble. We present an $SE(3)$-equivariant denoising-diffusion model that learns the configurational distribution of amorphous oxides, so the model itself is the structure database. The learning is data efficient. A model trained on $1{,}781$ DFT configurations suffices to reproduce partial radial distribution functions, coordination statistics and bond-angle distributions, and to generate models of over $3\times10^{5}$ atoms at a cost comparable to that of the cheapest classical pair potentials. The trained model can propose amorphous atomic structures for first-principles relaxation to explore the configuration space. For example, it locates an amorphous \ce{Zr}-\ce{Ta}-\ce{O} structure $36$~meV/atom below the previously known minimum. Generation can also extend beyond trained conditions to non-stoichiometric compositions, other mass densities, interfaces, and doping. First-principles verification confirms that generation can be steered to a requested energy, and shows that the denoising training loss does not rank generative quality, because the two measure different things.
\end{abstract}

\section{Introduction}
\label{sec:intro}

Atomic-scale modelling of crystalline and molecular systems is well established: classical pair and many-body potentials, DFT, and machine-learning interatomic potentials (MLIPs) such as NequIP~\cite{nequip}, MACE~\cite{mace} and Allegro~\cite{allegro}, together with periodic-table-wide models such as ALIGNN-FF~\cite{choudhary2023alignnff}, give accurate energies, forces and dynamics. Amorphous materials remain harder and more expensive, for three related reasons. They are not unique: they form a statistical ensemble whose realization depends on synthesis history, so no deterministic ``ground-truth'' structure can be tabulated and the standard practice of training against a few representative configurations is fundamentally undersampled. Amorphous databases are correspondingly immature beside the Inorganic Crystal Structure Database (ICSD) or the Materials Project for crystals, or PubChem and QM9 for molecules; only recently has a systematic computed amorphous database appeared~\cite{zheng2024amorphousdb}. And ``correct'' is preparation-dependent: one composition gives different short- and medium-range order under different cooling rates, pressures and dopant chemistries, and the amorphous phase shares much of its short-range order with its crystalline polymorphs~\cite{mavracic2018tio2,harrelson2021niobium}, making validation expensive and indirect.

All three problems share one origin: an amorphous material is defined by a distribution of structures, not by one structure. Diffusion models~\cite{sohl2015deep,ho2020denoising,song2021scoresde,dhariwal2021diffusion} are built to learn and sample such distributions. Trained on a finite set of DFT trajectories, a diffusion model represents the ensemble they were drawn from and can be sampled for any number of new configurations, so the trained model becomes the amorphous database: sampling it plays the role an ICSD lookup plays for crystals. Related work spans crystal-structure generators~\cite{xie2022cdvae,zeni2025mattergen,pakornchote2024dpcdvae}, surface-structure search~\cite{ronne2024surfacediff}, score dynamics~\cite{hsu2024scoredynamics}, latent diffusion transformers that generate crystalline and amorphous structures jointly~\cite{joshi2025allatom}, and amorphous-specific denoising networks~\cite{finkler2025amden,yang2025amorphousdiff,li2025amorphouscond,kwon2024spectroscopy}. Yang and Schwalbe-Koda~\cite{yang2025amorphousdiff} established diffusion models for amorphous materials and set a validation standard covering structure, mechanical properties and sampling diversity. ApolloX~\cite{li2025amorphouscond} addresses multi-element systems with a short-range-order descriptor, but builds its configurations by substituting species onto a fixed crystalline parent and relaxing to nearby minima; denoising from random coordinates instead samples the topologically disordered networks characteristic of glasses. Relative to the DM2 implementation~\cite{dm2,hsu2024score}, diffusion models add controllable conditioning of the generated ensemble.

This sampling problem limits applications of amorphous oxides in which atomic structure controls device performance. Two set the context. The first, which motivates this work, is the LIGO mirror coating problem: the amorphous oxide films used for the high-reflectivity coatings of gravitational-wave interferometer mirrors are annealed for tens to hundreds of hours at several hundred kelvin to reduce optical absorption and mechanical loss, over relaxation times molecular dynamics cannot reach~\cite{harry2007titania}. The \ce{TiO2}-doped \ce{GeO2} system studied here is the leading candidate, combining low mechanical loss with a refractive index high enough for a high-reflectivity stack~\cite{vajente2021tigeo2}. The second is decoherence in superconducting qubits, where two-level-system (TLS) defects in amorphous oxide interfaces~\cite{harrelson2021niobium} limit gate fidelity and coherence time; predicting TLS densities and energies from first principles needs converged sampling of rare motifs that small DFT supercells cannot provide. A generative model neither simulates the annealing trajectory nor enumerates defects one at a time: it samples directly from the configurational distribution in its training data, which is what makes both problems tractable. For the coating problem this gives a specific target, since mechanical loss depends on how far the film has been annealed and the useful window is bounded above by crystallization---too little annealing leaves the loss high, too much destroys the film. A model that reaches the low-energy amorphous states beyond the molecular-dynamics timescale, and exposes the underlying structures, would make that window a design variable.

This paper makes four contributions. The first is the model: a denoising-diffusion (DiT) model for amorphous oxides built on an $SE(3)$-equivariant message-passing network~\cite{nequip,graphite}, which reproduces DFT-quality partial radial distribution functions and coordination statistics from a few hundred atoms up to $3\times10^{5}$ (\cref{sec:genverify}). The second concerns how much data the ensemble takes to learn. A model trained on $1{,}781$ configurations, one fifth of the comparison set and drawn from $430$ DFT trajectories across twenty chemistries, reproduces the same structural observables at a modest cost in conditioning accuracy ($15.2$ against $11.6$~meV/atom mean absolute miss). What the model needs is coverage of the ensemble, not exhaustive sampling of it, and the same conclusion appears again in \cref{sec:conditioning}, where the accuracy with which a chemistry can be steered follows the range of energies its training labels span and not the size of the network. The third is that the learned ensemble extends past the conditions it was trained at: to non-stoichiometric compositions, to requested mass densities, to oxide--oxide interfaces and to doped compositions (\cref{sec:versatility}), and, followed by a single DFT relaxation, to an amorphous minimum below any previously reported (\cref{sec:explore}). The model is therefore a tool for structure search and not only for reproduction. The fourth is methodological and applies beyond this architecture: the denoising objective does not rank generative quality, because it is dominated by a noise range in which all the architectures we built are equivalent and is nearly blind to the narrow range that sets bond lengths and coordination (\cref{sec:lossquality}). Generative models of amorphous structure have to be selected on what they generate, verified by first-principles calculations or machine-learning potentials.


\section{Results}
\label{sec:results}

\subsection{Training set}
\label{sec:trainset}

Four related datasets appear in this paper, and which one a model saw matters, because the model-to-model comparisons of \cref{sec:lossquality} and the structures reported in \cref{sec:genverify,sec:explore,sec:versatility} do not come from the same weights.

Every comparison between models is made on one common set of $9{,}588$ configurations containing $5.16\times10^{6}$ atoms. Grouped by element set it holds twenty chemistries: the binary oxides \ce{GeO2}, \ce{TiO2}, \ce{SiO2}, \ce{HfO2}, \ce{Ta2O5}, \ce{ZrO2}, \ce{NbO2} and \ce{FeO}; elemental \ce{Si}; and eleven mixtures, of which \ce{Ti}-\ce{Ge}-\ce{O} ($2{,}349$ configurations) and \ce{Si}-\ce{Ti}-\ce{Ge}-\ce{O} ($2{,}002$) are much the largest, the remainder running from \ce{Zr}-\ce{Ta}-\ce{O} at $724$ down to \ce{Hf}-\ce{Si}-\ce{O} at $10$. Each chemistry contributes relaxed zero-temperature structures, snapshots from finite-temperature \emph{ab initio} molecular-dynamics trajectories, and configurations compressed or expanded away from equilibrium volume.

The other three sets are an earlier state of the same collection holding $8{,}882$ configurations, which is what the generator behind the structures of \cref{sec:genverify,sec:explore,sec:versatility} was trained on; the deliberately sparse set of $1{,}781$ configurations described below; and a $15{,}826$-configuration expansion used for the corpus-expansion arm of \cref{tab:checkpoints}. The interface and multilayer cells of \cref{sec:versatility} come from an earlier checkpoint again, of $593{,}392$ parameters. We name the set behind every reported structure and every model ranking; no comparison in this paper crosses two of them.
\Cref{tab:datasets} lists the four datasets and the result each one is behind.

\begin{table}[t]
\centering
\caption{The four datasets, and which result each one is behind. ``Configurations'' counts DFT structures after frame selection. The comparison set holds $5.16\times10^{6}$ atoms. Both the full-corpus and sparse-corpus models have $606{,}584$ parameters; the interface and multilayer cells of \cref{sec:versatility} come from an earlier checkpoint of $593{,}392$.}
\label{tab:datasets}
\small
\begin{tabular}{@{}lrp{0.40\linewidth}l@{}}
\toprule
dataset & configs. & used for & model \\
\midrule
comparison set  & $9{,}588$  & model-to-model ranking (\cref{sec:lossquality}) & various \\
full corpus     & $8{,}882$  & structures of \cref{sec:genverify,sec:explore,sec:versatility} & full-corpus \\
sparse set      & $1{,}781$  & data-requirement test (\cref{tab:liteverify}, \cref{fig:ensemble}) & sparse-corpus \\
expanded corpus & $15{,}826$ & corpus-expansion arm of \cref{tab:checkpoints} & --- \\
\bottomrule
\end{tabular}
\end{table}

The cation distribution is strongly non-uniform. Of $1.74\times10^{6}$ cations, \ce{Ge} accounts for $49.8\%$ and \ce{Ti} for $27.2\%$, followed by \ce{Si} ($8.0\%$), \ce{Ta} ($6.1\%$), \ce{Hf} ($4.7\%$), \ce{Zr} ($3.7\%$) and \ce{Nb} ($0.52\%$). \Cref{sec:conditioning} shows that this imbalance, together with the range of energies sampled within each family, sets how accurately the model can be steered. The one chemistry it cannot generate is the one whose cation is scarcest.

The sparse dataset exists to test how much data learning the ensemble actually requires. It holds $1{,}781$ configurations taken from $430$ DFT trajectories, spanning the same twenty chemistries, at up to six frames per trajectory. It is one fifth the size of the comparison set, and no family contributes more than $49$ trajectories; amorphous \ce{GeO2}, the chemistry most of the tests in this paper use, is represented by $28$. A model trained on it is referred to below as the sparse-corpus model. It is otherwise identical to the full-corpus model: same architecture, same $606{,}584$ parameters, same schedule.

\subsection{Model generation and atomic structure verification}
\label{sec:genverify}
\label{sec:scaling}
\label{sec:rdf}

\begin{figure*}[t]
 \centering
 \begin{subfigure}[b]{0.45\linewidth}
 \includegraphics[width=\linewidth]{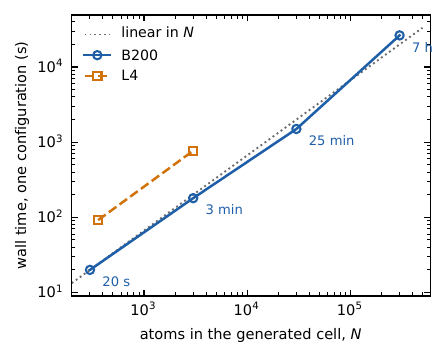}
 \caption{Generation cost against system size}
 \end{subfigure}\hfill
 \begin{subfigure}[b]{0.38\linewidth}
 \includegraphics[width=\linewidth]{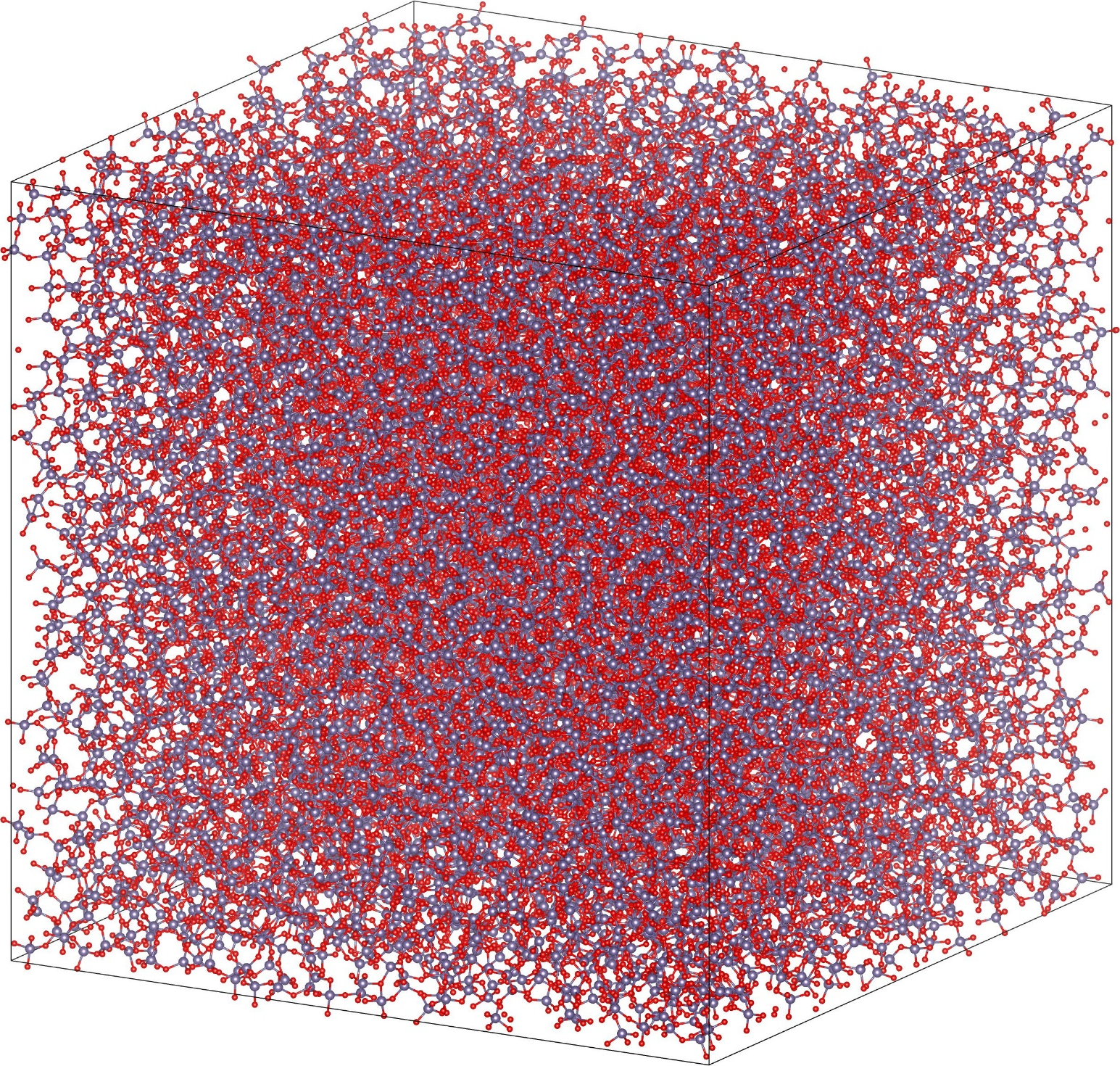}
 \caption{The generated $N{=}300\,000$-atom a-\ce{GeO2} cell}
 \end{subfigure}

 \vspace{0.5em}
 \begin{subfigure}[b]{0.99\linewidth}
 \includegraphics[width=\linewidth]{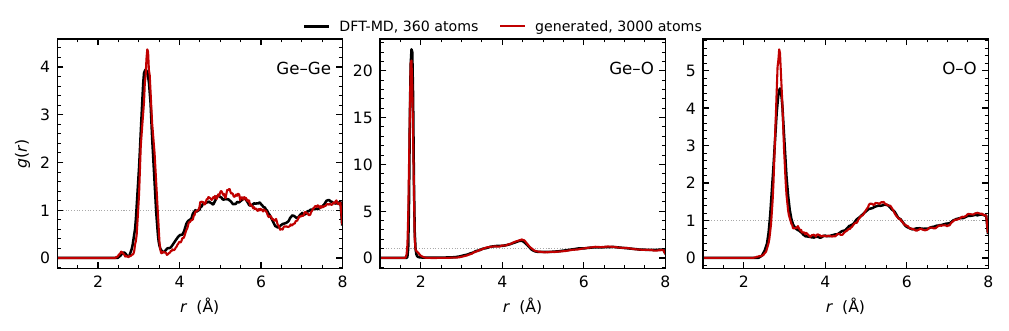}
 \caption{Partial RDFs of the generated $N{=}3000$-atom cell against DFT-MD}
 \end{subfigure}
 \caption{Generation at scale and its structural verification. \textbf{a},~Wall time for one generated configuration against atom count, with almost linear scaling; circles a single B200, squares a single L4. \textbf{b},~The generated $N{=}300\,000$-atom a-\ce{GeO2} cell. \textbf{c},~The three partial RDFs (Ge--Ge, Ge--O, O--O) of a generated $N{=}3000$-atom cell (red)} against a $360$-atom DFT-MD \emph{trajectory average} over $1000$ frames spanning \SI{1}{\pico\second}, ramped from $10$ to $300$~K (black). Consecutive frames are correlated, so this is one trajectory average and not $1000$ independent samples. The RDFs are matching perfectly except slightly broadening due to temperature effect of DFT-MD.
 \label{fig:genverify}
 \label{fig:scaling}
 \label{fig:rdf-geo2}
\end{figure*}

Each generation run produces an independent configuration rather than a correlated frame of a trajectory, and its cost is approximately linear in the number of atoms, limited in practice by GPU memory rather than by computation. \Cref{fig:genverify}(a) shows representative wall times on a single NVIDIA B200 GPU: \SI{20}{\second} for $N{=}300$ atoms, \SI{180}{\second} for $N{=}3000$, \SI{1500}{\second} ($\sim$25 min) for $N{=}30\,000$, and \SI{26400}{\second} ($\sim$7~h) for $N{=}300\,000$. The $N{=}300\,000$ configuration is a \ce{GeO2} cube about $17$~nm on a side, and is the largest amorphous oxide configuration we are aware of from a generative model. Nothing in the network fixes the system size: the same weights can generate $300$, $300\,000$ atoms or other, because the model predicts a per-atom displacement from a local neighbourhood and the cell is enlarged, not retrained for.

This is the range that makes the sampling problem tractable. DFT-MD is not feasible at $N=300\,000$ atoms, requiring many orders of magnitude more compute than is available; the $1008$-atom \ce{Ti}-\ce{Ge}-\ce{O} configuration used below took \SI{278}{\second} on one L4 GPU against roughly two weeks of cluster time for the $720$-atom DFT reference calculation it is compared with. Classical pair potentials remain faster per configuration, by a factor of $2$--$5$ for Buckingham-type forms. For sampling an ensemble, however, the relevant cost is per independent configuration rather than per frame: each diffusion run returns one independent sample from the target distribution, whereas an MD trajectory yields one independent sample per decorrelation time.

Generated configurations are verified against DFT references on two observables: the partial radial distribution functions (RDFs), which test pair correlations out to the medium-range scale, and the coordination-number (CN) distribution, which tests the network topology that the RDF only implies. Both are needed, because an RDF averages over the cell and can look correct while the coordination statistics are wrong: it is the coordination distribution that separates a tetrahedral glass from a network with the same mean bond length and too many five-fold sites.

\Cref{fig:genverify}(c) compares the three partial RDFs (Ge--Ge, Ge--O, O--O) of a generated \ce{GeO2} cell with DFT molecular dynamics, as a trajectory average over $1000$ consecutive frames of $360$ atoms along a single \SI{1}{\pico\second} run in which the temperature is ramped from $10$ to $300$~K (mean $155$~K); the frames are correlated and the average is not over $1000$ independent configurations. The generated cell is the $N{=}3000$ one, because the coordinates of the $N{=}300\,000$ run were not retained; that run is quoted for generation cost and size only.

Coordination is reproduced as closely as the pair correlations. With both sides passed through one analysis path, the medium-range Ge--Ge peak near \SI{3.2}{\angstrom}, which corresponds to corner-sharing tetrahedra, is reproduced together with the shoulder produced by edge-sharing defects. The same path puts the Ge--O first peak at \SI{1.770}{\angstrom} against \SI{1.770}{\angstrom}, O--O at $2.870$ against \SI{2.890}{\angstrom} and Ge--Ge at $3.210$ against \SI{3.170}{\angstrom}; the mean Ge coordination is $4.050$ against $4.042$ and the mean O coordination $2.025$ against $2.021$, with $95\%$ of Ge sites four-fold and a smallest interatomic separation of \SI{1.648}{\angstrom}. The tetrahedral network of \ce{GeO2} is therefore reproduced as a topology and not only as a first-peak position.

For a multi-element test we trained on \ce{TiO2}-doped \ce{GeO2} and compared against a $720$-atom DFT reference at 44\% \ce{TiO2}, obtained by ionic relaxation of a classically quenched configuration and averaged over the last ten ionic steps. We measured all six partial RDFs, through one code path for both sides, for a generated $1008$-atom \ce{Ti148Ge188O672} configuration at \SI{3.700}{\gram\per\cubic\centi\meter}, the same composition and density as the reference. Five of the six first-neighbour peaks agree to within \SI{0.02}{\angstrom}: Ti--O $1.830$ against \SI{1.830}{\angstrom}, Ge--O $1.750$ against $1.770$, O--O $2.850$ against $2.870$, Ge--Ge $3.170$ against $3.190$ and Ti--Ge $3.470$ against \SI{3.510}{\angstrom}. The exception is Ti--Ti, $3.590$ against \SI{3.450}{\angstrom}: the rarest pair, sampled by only $148$ atoms here and $105$ per frame in the reference, and the correlation neither calculation converges well. Its seed-to-seed scatter shows this directly---across three independently generated cells the Ti--Ti first peak moves by \SI{0.32}{\angstrom} ($3.31$, $3.59$, \SI{3.63}{\angstrom}) while Ti--Ge moves by \SI{0.04}{\angstrom}---so the \SI{0.14}{\angstrom} discrepancy lies well inside that pair's sampling spread. Coordination is reproduced with the pair correlations---\ce{Ti} $5.115$ against $5.089$, \ce{Ge} $4.160$ against $4.037$, \ce{O} $2.290$ against $2.249$---so \ce{Ge} stays four-fold while \ce{Ti} takes the higher coordination it has in \ce{TiO2}: a single homogeneous network hosting two cation polyhedra, not two segregated regions.

Both comparisons above use the full-corpus model. The same observables were then measured for the sparse-corpus model of \cref{sec:trainset}, the test of how much data learning this ensemble requires. We generated three independent configurations of each of three chemistries at the density of a DFT reference of the same composition, and computed partial RDFs and coordination distributions of both through one code path (\cref{tab:liteverify}). Every cation--oxygen first-peak position is reproduced to within \SI{0.01}{\angstrom} and every mean coordination number to within $0.07$. For \ce{Ti_{0.44}Ge_{0.56}O2} and \ce{TiO2} the reference trajectory is not among the $1{,}781$ the model saw, so those two test generalization rather than recall. The comparison uses DFT ionic relaxations at the same density rather than a finite-temperature average, because the generator runs at zero effective temperature and returns an inherent structure, so comparing its peak \emph{widths} against a thermally broadened average would not be meaningful. Generation is also size-independent on these observables: at $3000$ atoms the same model gives a Ge coordination of $4.035$ against $4.053\pm0.010$ at $360$ atoms, with an identical Ge--O first peak.

\begin{figure}[t]
 \centering
\color{blue}
 \includegraphics[width=0.99\linewidth]{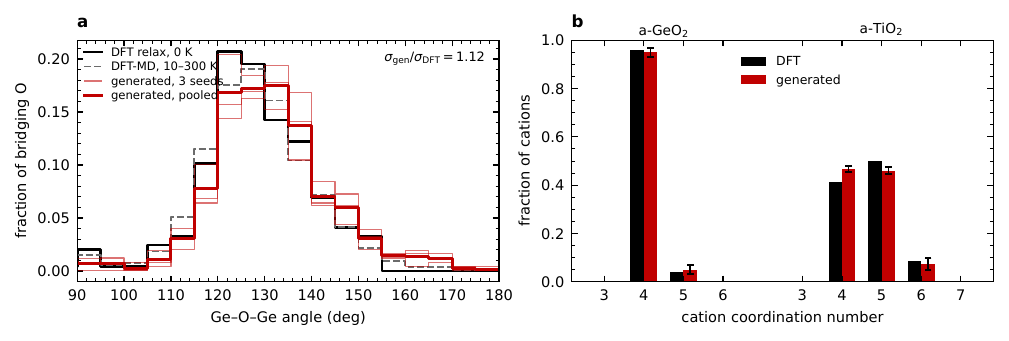}
 \caption{The generated ensemble reproduces distribution \emph{widths}, not only means. \textbf{a},~\ce{Ge}--\ce{O}--\ce{Ge} bond angle distribution of a-\ce{GeO2} at \SI{3.75}{\gram\per\cubic\centi\meter}: three independently generated cells (thin red), their pooled distribution (thick red), the $0$~K DFT ionic relaxation (black) and the $10$--$300$~K DFT-MD trajectory average (grey dashed). The generated distribution is $1.12$ times as wide as the DFT one; the generated-to-DFT total-variation distance, $0.119$, is smaller than the $0.133$ between two generated cells. \textbf{b},~Full cation coordination-number distribution, for a-\ce{GeO2} and a-\ce{TiO2}: DFT ionic relaxation (black) against the mean over three generated cells (red), error bars the spread across those three.}
 \label{fig:ensemble}
\end{figure}

Everything above compares first moments (peak positions and mean coordination numbers) which is also what a mode-collapsed generator could produce. The claim that the model learns the \emph{ensemble} requires the width of the generated distribution, so we measured two distributional observables against the same references through one code path (\cref{fig:ensemble}). The \ce{Ge}--\ce{O}--\ce{Ge} bridging angle of a-\ce{GeO2} at \SI{3.75}{\gram\per\cubic\centi\meter} is a continuous order parameter, so its width is meaningful. Pooling three independently generated cells gives $131.2^\circ$ with a standard deviation of $13.1^\circ$, against $127.9^\circ$ and $11.6^\circ$ for the $0$~K DFT relaxation and $128.1^\circ$ and $12.2^\circ$ for the $10$--$300$~K DFT-MD trajectory: the generated distribution is $1.12$ times \emph{as wide} as the DFT one, the opposite of mode collapse. On a common $5^\circ$ grid the total-variation distance between the pooled generated and DFT relaxed distributions is $0.119$, smaller than the $0.133$ between two generated cells, so the ensemble differs from the reference by less than its own sampling noise. Second, the full coordination histogram, not its mean, is reproduced across four chemistries measured at seven densities: over the $15$ (chemistry, element) pairs of a-\ce{GeO2}, a-\ce{TiO2}, a-\ce{Ti_{0.44}Ge_{0.56}O2} and a-\ce{HfO2} the median total-variation distance is $0.043$ against $0.033$ between generated replicas, and $12$ of $15$ agree to within $0.10$. The generator reproduces the defect populations that give those distributions their width---for a-\ce{GeO2} placing $4$--$8\%$ of \ce{Ge} in five-fold sites against $4.2\%$ in the DFT relaxation---rather than collapsing onto the ideal four-fold motif.

\begin{table}[t]
\centering
{%
\caption{Structural verification of the sparse-corpus model, trained on $1{,}781$ configurations. Three independently generated configurations per chemistry, compared with a DFT reference of the same composition at the same mass density; $r_1$ is the first-peak position of the partial RDF and CN the mean coordination number of the cation, using first-shell cutoffs of \SI{2.40}{\angstrom} (Ge--O) and \SI{2.60}{\angstrom} (Ti--O). Uncertainties on generated quantities are the spread over the three configurations, and $d_{\min}$ is the smallest interatomic distance in the generated cells, against a \SI{1.45}{\angstrom} chemistry limit.}
\label{tab:liteverify}
\footnotesize
\setlength{\tabcolsep}{5pt}
\begin{tabular}{lcccccc}
\toprule
system & atoms & $\rho$ (g\,cm$^{-3}$) & pair & $r_1$ gen / DFT (\AA) & CN gen / DFT & $d_{\min}$ (\AA) \\
\midrule
a-\ce{GeO2}           & $360$ & $3.746$ & Ge--O & $1.770$ / $1.770$ & $4.050\pm0.022$ / $4.042$ & $1.707$ \\
a-\ce{Ti_{0.44}Ge_{0.56}O2} & $300$ & $3.700$ & Ti--O & $1.837$ / $1.830$ & $5.045\pm0.023$ / $5.089$ & $1.693$ \\
                      &       &         & Ge--O & $1.770$ / $1.770$ & $4.107\pm0.036$ / $4.037$ &         \\
a-\ce{TiO2}           & $300$ & $2.660$ & Ti--O & $1.830$ / $1.830$ & $4.607\pm0.045$ / $4.675$ & $1.739$ \\
\bottomrule
\end{tabular}}
\end{table}

\subsection{Model selection: training loss versus generation quality}
\label{sec:lossquality}

A diffusion model is trained to predict the noise added to an atomic configuration, so the natural measure of accuracy is the mean squared error of that prediction on held-out data. Ranking architectures by this validation error is standard practice. For amorphous structure generation it does not necessarily give the right order, and the reason is the noise scale.

The objective of \cref{eq:loss} is an expectation over $\sigma\in[10^{-3},0.75]$~\AA{}, and the variance of the target is itself $\sigma^{2}$, so the three decades of $\sigma$ do not contribute equally. Measured on our own loss curves, $84\%$ of the loss sits at $\sigma>0.5$~\AA{}, where three architectures spanning $447$k to $651$k parameters agree to within $3.6$ percentage points. Only $4.4\%$ sits in $\sigma\in[0.02,0.35]$~\AA{}, the band in which bond lengths and coordination polyhedra are established, and there the same architectures differ by up to a factor of two. At the bottom of the range the denoiser has no skill at all: at $\sigma=0.005$~\AA{} its error is $6.7\times$ worse than predicting zero displacement, because the injected noise is far below the intrinsic first-shell spread of \SI{0.155}{\angstrom}. The scalar loss is therefore a precise measurement of the regime where all models agree and a weak one of the regime where they differ.

The divergence is bounded by how far the two can go apart. The variant reaching the lowest validation error of any model we have trained on the unweighted loss function \cref{eq:loss}, $0.085$, generates no usable structure at any requested energy or density: all $100$ configurations in a test set contained fused atoms, with DFT energies $3.5$ to $16.6$~eV/atom above the amorphous minimum. An independently trained model of the same design, again the lowest-error member of its cohort, produced $61$ of $100$ fused. The failure reproduces in two separate implementations of the sampler, so it lies in the weights. That sample quality and average likelihood can diverge is known in generative modelling~\cite{theis2016note}; here the divergence is unambiguous, because a failed sample is geometrically impossible rather than subjectively poor. These variants add a long-range branch to the network; they reach the lowest training error we have measured and give no improvement in generation, so we treat them as future work and every configuration in this paper was generated with the short-range backbone of \cref{sec:arch:graphite}.

Training a single model for longer behaves the same way. \Cref{tab:checkpoints} follows one run at three checkpoints. Its validation loss falls monotonically, $0.28213\to0.27393\to0.27290$, while its ability to hit a requested energy peaks in the middle and then decays: the calibration slope crosses unity from above and keeps falling, $2.039\to1.170\to0.676$, the mean absolute energy miss over $204$ DFT-verified configurations goes $46.1\to17.1\to18.6$~meV/atom, and the bias changes sign. Resolving the same validation loss by noise scale explains it. The bin $\sigma\in[0,0.1]$~\AA{}, which contains the $\sigma=0.020$~\AA{} at which generation performs its final refinement, holds $0.026\%$ of the objective and did not move at all between the last two checkpoints ($0.00022$ against $0.00022$), whereas the bins above $\sigma=0.6$~\AA{} hold $72.8\%$ of the loss and supplied $51\%$ of the total improvement. The loss went down in the range generation never visits.

\begin{table}[t]
\centering
{%
\caption{One training run at three checkpoints: validation loss against DFT-verified conditioning accuracy. The scan is $51$ requested excess energies from $0$ to $0.100$~eV/atom in $0.002$ steps, four repeats, \ce{Ge120O240} at \SI{3.650}{\gram\per\cubic\centi\meter}, one DFT single point per configuration ($n=204$ per checkpoint). $|\Delta E$ miss$|$ is the mean absolute difference between achieved and requested excess energy; ``repeat $\sigma$'' is the spread across repeats at a fixed request, that is the error bar on one generated structure; $R^{2}$ is for achieved against requested. Losses here are on the $1/\sigma^{2}$-weighted objective (\cref{sec:setup}), with $25\%$ of each batch held at $\sigma=10^{-3}$ with an exact force label. They are comparable down the column, not with losses quoted elsewhere in this paper.}\label{tab:checkpoints}}
\begin{tabular}{lcccccc}
\toprule
steps & valid loss & slope & $|\Delta E$ miss$|$ & bias $\pm$ SE & repeat $\sigma$ & $R^{2}$ \\
 & & & (meV/atom) & (meV/atom) & (meV/atom) & \\
\midrule
$560$k          & $0.28213$ & $2.039\pm0.101$ & $46.1$ & $+38.7\pm4.2$ & $22.7$ & $0.805$ \\
$2.48$M         & $0.27393$ & $1.170\pm0.034$ & $17.1$ & $+13.8\pm1.1$ & $11.2$ & $0.852$ \\
$4.04$M         & $0.27290$ & $0.676\pm0.052$ & $18.6$ & $-8.8\pm1.7$  & $19.5$ & $0.456$ \\
\bottomrule
\end{tabular}
\end{table}

The model does not stop responding to the request. We measured the response directly, as $S_E = \lVert \hat{\bm{\varepsilon}}(\Delta E{=}0.10) - \hat{\bm{\varepsilon}}(\Delta E{=}0)\rVert / \lVert \hat{\bm{\varepsilon}}(\Delta E{=}0)\rVert$ on twelve fixed configurations at $\sigma=0.020$~\AA{}, the same twelve for every checkpoint, so that the probe cannot be confounded by the structures a checkpoint happens to generate. Across $3.5$M steps $S_E$ fell by $8\%$ ($1.07\to1.03\to0.98$) while the norms of the weights carrying the energy request \emph{grew} by $44\%$ and $89\%$. Over the same interval the realized slope fell threefold and $R^{2}$ halved. The response is as large as it ever was and is no longer aligned with the energy axis, which is a different failure from a weak response and would not be repaired by strengthening the conditioning.

Two further properties make the reported loss unsafe to compare between runs. It is not comparable across datasets: on the $1/\sigma^{2}$-weighted objective the sparse-corpus model of \cref{sec:trainset} reaches the lowest validation loss of that cohort, $0.29580$ against $0.32815$ for the full-corpus model, because one fifth as much data is easier to denoise. That value and the $0.085$ above are on the two different objectives of \cref{sec:setup}, so they are not comparable with each other. And the loss depends on settings that leave the weights untouched, since it is a mean over minibatch means and the per-GPU batch size therefore re-weights configurations of different atom count: on the unweighted objective one fixed checkpoint, evaluated at two batch sizes at an identical step, returned $0.0831$ and $0.0885$.

We therefore select models on what they generate. Configurations are produced over a range of requested conditions, screened on a minimum interatomic distance set by the chemistry (\SI{1.45}{\angstrom}) rather than by the sampler's \SI{1.1}{\angstrom} steric floor, and evaluated with DFT. This test does not depend on how the training run was configured, and it is the criterion used in \cref{sec:explore,sec:conditioning}.

\subsection{DFT-assisted exploration of the amorphous energy landscape}
\label{sec:explore}
\label{sec:assistedopt}
\label{sec:newminima}

The trained model can be used to propose configurations for subsequent DFT relaxation, and this turns generation into a search. Generated \ce{GeO2} configurations, each followed by a single DFT relaxation, frequently relax to energies below the lowest energy in the training set: for models trained on a single relaxation trajectory plus a single $300$~K MD trajectory, whose labels span $[-6.3414,-6.2980]$~eV/atom, the relaxed configurations reach $\approx-6.41$~eV/atom, some $70$~meV/atom below the training minimum. The model therefore does not only reproduce its training distribution. An information-theoretic analysis of comparable atomistic training sets~\cite{schwalbekoda2025information} shows that the entropy of a typical DFT-MD ensemble leaves room for configurations of this kind. The procedure is equivalent to enhanced-sampling MD with the diffusion model in place of a bias potential, at a small fraction of the cost.

Interpreting such a number requires two reference energies. First, $-6.41$~eV/atom must not be read as the onset of crystallization. We obtain $-6.4133$~eV/atom for $\alpha$-quartz \ce{GeO2} and $-6.4141$~eV/atom for the best amorphous structure: at the PBE level the two are degenerate to $0.8$~meV/atom, with the glass marginally lower, and the rutile-structured polymorph lies $5.8$~meV/atom \emph{above} the glass, a known functional artifact. Reaching $-6.41$~eV/atom is what reaching the amorphous inherent-structure minimum looks like, not the onset of order. In \ce{GeO2} the amorphous-to-crystalline boundary therefore cannot be located from energy at all; it requires structural order parameters. Second, the lowest energy in a training set is an unsuitable reference: it records how much sampling that dataset received, not a property of the material. We therefore fix per family the relaxed crystal energy $E_{\mathrm{xtl}}$ and the lowest known amorphous energy $E_{\mathrm{am}}$, measuring discovery against the latter. The two are mutually consistent, the amorphization energy $E_{\mathrm{am}}-E_{\mathrm{xtl}}$ falling in the narrow range $+0.09$ to $+0.16$~eV/atom across every oxide studied.

With both references fixed the procedure is a search. We condition the model at a target energy at or slightly below the lowest known amorphous energy of a family, sweep the requested mass density across the plausible range for that composition, and relax each generated configuration with DFT. The cost is set by the relaxations, not by the generation. Applied across the oxide families of the training set, this located a configuration below the previously known amorphous minimum in \ce{Zr}-\ce{Ta}-\ce{O}: $-9.5713$~eV/atom against a previous best of $-9.5356$, a gain of $35.7$~meV/atom, at a mass density of \SI{6.50}{\gram\per\cubic\centi\meter} and an amorphization energy of $+0.110$~eV/atom above the relaxed crystal. Both energies are for fully relaxed inherent structures obtained under one common protocol, so the comparison is minimum to minimum. The record cell is \ce{Ta40Zr40O180} ($260$ atoms), that is \ce{Zr}:\ce{Ta}${=}1{:}1$ by cation or $67$~mol\% \ce{ZrO2}---a higher zirconia fraction than the $30\%$ \ce{ZrO2}-doped \ce{Ta2O5} coating cell of \cref{sec:versatility}.

The gain is small in absolute terms but it is reproducible: the record was obtained independently from distinct generated starting configurations and relaxed to a maximum residual force below $0.05$~eV/\AA{}. That tolerance bounds the energy still to be gained: for a harmonic basin the residual energy is $\sum_i|\bm{F}_i|^{2}/2k$, which for $|\bm{F}|_{\max}=0.05$~eV/\AA{} and a conservative oxide force constant $k=5$~eV/\AA{}$^{2}$ is at most $0.25$~meV/atom, two orders of magnitude below the gain reported above. The protocol control is complete: we re-relaxed the previous record under the identical protocol and recovered it to within $1$~meV/atom, excluding a difference in method as the source of the improvement.
The baseline is the lowest amorphous energy of that family known before this search began, that is the floor carried in the conditioning labels of the generating checkpoint, and not a literature value; it was then lowered in several independent rounds of the same generate--relax loop, through five intermediate values, so the intermediates are products of the method being reported and the pre-search floor is the only baseline independent of it. The residual force bound above is per atom, so it does not grow with cell size.

A trained model therefore acts as a database that can be sampled for structures which were never computed to build it, and the loop closes on itself: each new record lowers the reference energy used to label the next round of training data. The division of labour is the useful part. The model supplies the configurational search, which is the expensive step for an amorphous material because the basins are numerous and unrelated to one another, and DFT supplies the energy and the final relaxation, which is what makes the result quantitative.

\subsection{Generalization and generation beyond training}
\label{sec:versatility}

Nothing in the sampler requires the requested configuration to resemble a training configuration, and the useful question is how far outside the training conditions the learned ensemble remains physical. We tested four kinds of departure: stoichiometry, mass density, interfaces between two different oxides, and doping. In each case the output is a complete periodic atomic model, which is what a subsequent DFT, MLIP or transport calculation needs as a starting point.

\begin{figure*}[tp]
 \small
 \centering
 \begin{subfigure}[t]{0.24\linewidth}
 \includegraphics[width=\linewidth]{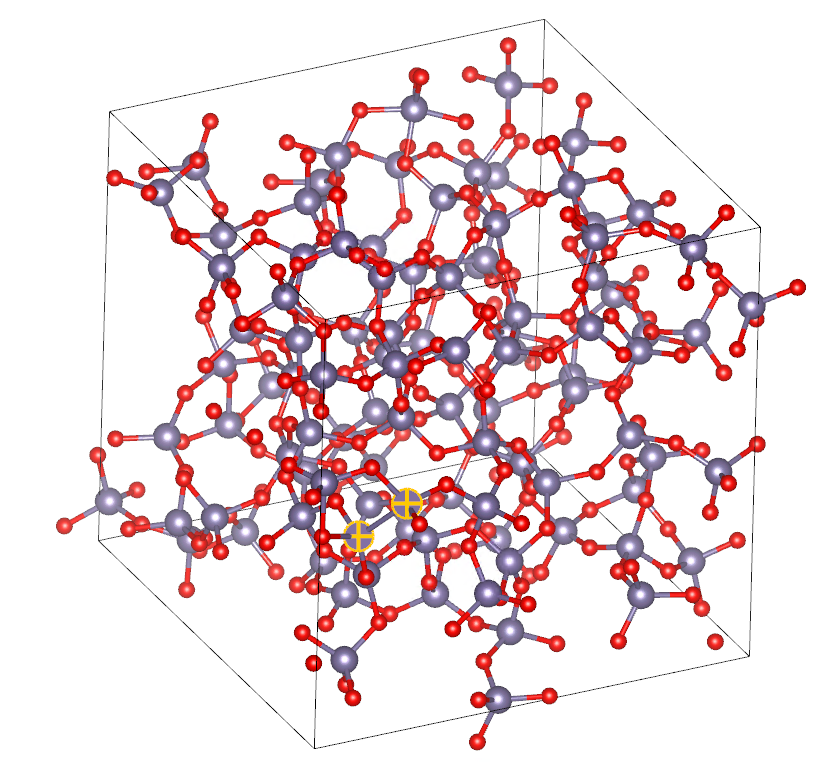}
 \caption{Non-stoichiometric a-\ce{Ge100O199}}
 \end{subfigure}\hfill
 \begin{subfigure}[t]{0.24\linewidth}
 \includegraphics[width=\linewidth]{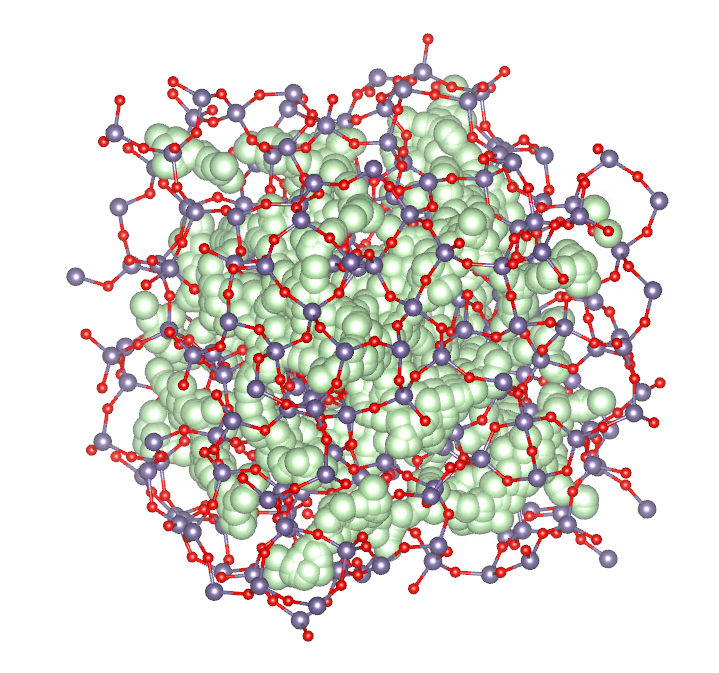}
 \caption{Low-density a-\ce{GeO2}, $\rho{=}3.0$~g/cm$^3$}
 \end{subfigure}\hfill
 \begin{subfigure}[t]{0.33\linewidth}
 \includegraphics[width=\linewidth]{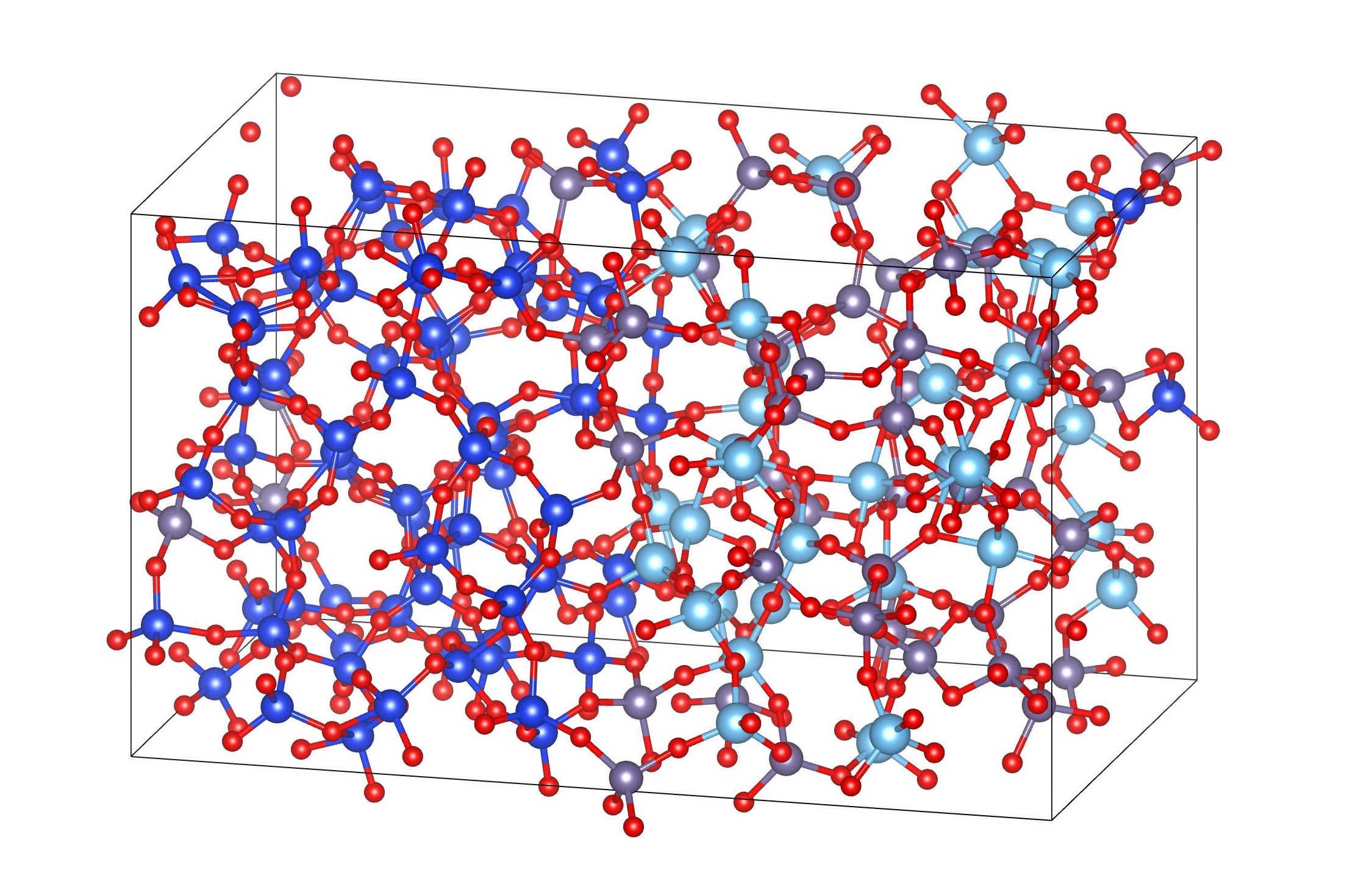}
 \caption{a-\ce{SiO2}/a-(\ce{Ti},\ce{Ge})\ce{O2} interface, $480$ atoms}
 \end{subfigure}
  \begin{subfigure}[t]{0.58\linewidth}
 \includegraphics[width=\linewidth]{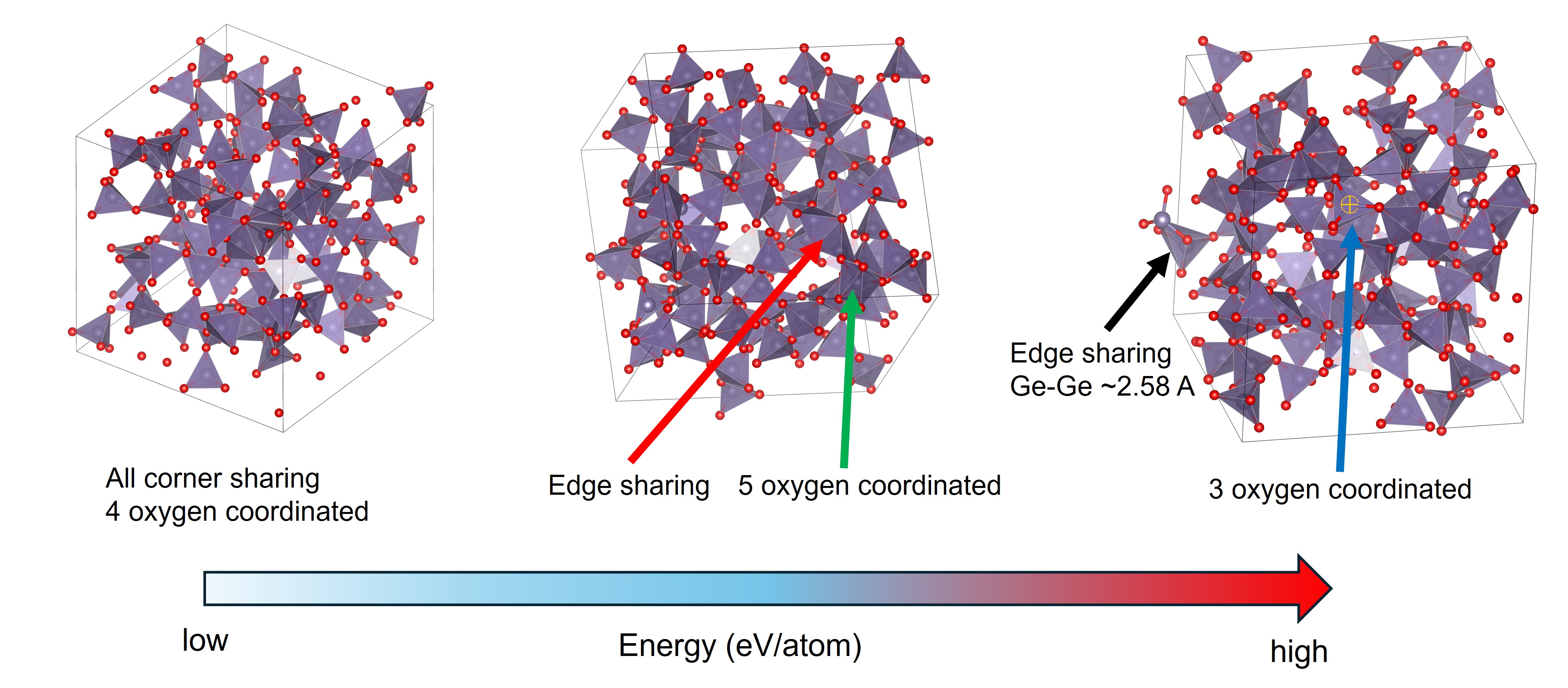}
 \caption{Energy vs atomic structure of a-\ce{GeO2}}
 \end{subfigure}
 \begin{subfigure}[t]{0.3\linewidth}
 \includegraphics[width=\linewidth]{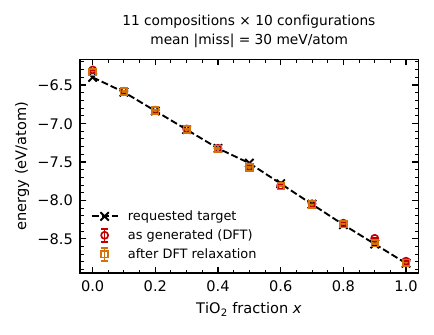}
 \caption{(\ce{Ti}$_x$\ce{Ge}$_{1-x}$)\ce{O2} doping series}
 \end{subfigure}
 \caption{Generation beyond the training conditions: composition, density, interfaces and doping. \textbf{a},~\ce{Ge100O199}, a non-stoichiometric composition. \textbf{b},~\ce{GeO2} at a requested mass density of \SI{3.0}{\gram\per\cubic\centi\meter}; green spheres mark void space. \textbf{c},~A generated a-\ce{SiO2}/a-(\ce{Ti},\ce{Ge})\ce{O2} interface, $480$ atoms in a $16\times16\times27$~\AA{} cell with the two layers at their own densities and the stacking axis horizontal; \ce{O} red, \ce{Si} dark blue in the left-hand region, the two cations of the (\ce{Ti},\ce{Ge})\ce{O2} region light blue and grey. \textbf{d},~Generated a-\ce{GeO2} configurations ordered by energy, with the motifs that appear as the energy rises: all-corner-sharing, four-coordinated \ce{Ge} at the low-energy end; edge-sharing tetrahedra and five-coordinated oxygen at intermediate energy; and three-coordinated oxygen with an edge-sharing \ce{Ge}--\ce{Ge} pair at \SI{2.58}{\angstrom} at the high-energy end. \textbf{e},~The (\ce{Ti}$_x$\ce{Ge}$_{1-x}$)\ce{O2} at $11$ compositions, ten configurations each: requested target (black crosses), DFT energy as generated (red circles) and after DFT relaxation (orange squares); error bars are the standard deviation over the ten.}
 \label{fig:versatile}
 \label{fig:interface}
\end{figure*}

\emph{Stoichiometry and density.} \Cref{fig:versatile}(a) is a non-stoichiometric \ce{Ge100O199} configuration generated at a target energy of $-6.34$~eV/atom, that is \ce{GeO2} with one oxygen vacancy; no configuration of this composition appears in the training set. Panel~(b) is \ce{GeO2} generated at a requested density of \SI{3.0}{\gram\per\cubic\centi\meter}, well below the equilibrium value, with the resulting void space shown in green. Panel~(d) shows the local structure that accompanies rising energy: mostly corner-sharing tetrahedra together with edge-sharing tetrahedra and a Ge--Ge separation of \SIrange{2.57}{2.58}{\angstrom}, and both three- and five-coordinated oxygen sites. All of these motifs are expected in an oxygen-deficient amorphous oxide and are reported in DFT studies of reduced germanate glasses. The failure mode at large density changes is also informative: when the requested density is far below equilibrium the model draws the corpus-average coordination rather than a density-appropriate one, so under-coordination rises with the size of the requested drop.

\emph{Interfaces.} An interface is a composition and a density that both vary with position, and it is the configuration that matters most for the two applications of \cref{sec:intro}. Supplying a target composition per region rather than per cell generates two-region and multilayer stacks directly, with no separate slab construction or welding step. \Cref{fig:versatile}(c) shows the simplest case at a size where the individual atoms are visible: one a-\ce{SiO2}/a-(\ce{Ti},\ce{Ge})\ce{O2} junction, $480$ atoms in a $16\times16\times27$~\AA{} cell with the two layers requested at their own densities of $2.23$ and \SI{3.70}{\gram\per\cubic\centi\meter}, generated in \SI{91}{\second} from a single global energy target of $-7.0$~eV/atom. Each cation keeps the local environment it has in its own oxide. The three cation--oxygen first shells sit at \SI{1.644}{\angstrom} (Si--O), $1.756$ (Ge--O) and \SI{1.831}{\angstrom} (Ti--O); \ce{Si} and \ce{Ge} are four-fold on their own sides, mean $4.13$ and $4.09$ with mode four, while \ce{Ti} spreads over four-, five- and six-fold sites at mean $4.60$; and the smallest interatomic distance is \SI{1.55}{\angstrom}.

The construction scales, and boundaries sharpen as layers thicken. The most demanding case is a four-layer stack of three oxides in one $36\,000$-atom cell, \ce{SiO2} $\vert$ \ce{GeO2} $\vert$ \ce{TiO2} $\vert$ \ce{SiO2} along $z$ at $2.23$, $3.70$, $3.90$ and \SI{2.23}{\gram\per\cubic\centi\meter}, containing three oxide--oxide interfaces built in one pass. In its $40\times40\times325$~\AA{} cell the layers are $168$, $90$ and \SI{68}{\angstrom} thick and the cation fraction goes from $10\%$ to $90\%$ over \SI{4.0}{\angstrom}, about one bond length. Cation--oxygen shells fall at the same separations as in the small cell, cations take the coordination of their own oxides, and every cross-layer cation pair is suppressed by one and a half to two orders of magnitude relative to within-layer pairs. Those cross-junction correlations are far larger in the $480$-atom cell, as they should be: a \SI{27}{\angstrom} cell is mostly interface, so the suppression measures layer thickness rather than mixing. The construction reaches a two-region \ce{SiO2}/(\ce{Ti_{0.44}Ge_{0.56}})\ce{O2} bilayer of $138\,345$ atoms by domain decomposition across several GPUs. Coordination survives a sharp boundary: in a $240$-atom a-\ce{SiO2}/a-\ce{GeO2} interface both \ce{Si} and \ce{Ge} average $4.03$ with mode four, so neither cation is forced out of tetrahedral coordination. Amorphous films on crystalline substrates behave the same way from $624$ to $26\,175$ atoms, under two conditions: the substrate must be terminated at a natural cleavage plane or the film does not bond to it, and the chemistries do not scale equally---a-\ce{TiO2} on rutile~(110) reached $26\,175$ atoms directly, whereas the tallest a-\ce{GeO2} film on $\alpha$-quartz~(0001) collapsed when generated in one pass and had to be built as narrow bands, each denoised onto the one below. For the gate-stack-like a-\ce{SiO2}/a-\ce{HfO2} system, six independent $336$-atom interfaces (\ce{Si52O104} $\vert$ \ce{Hf60O120}, two periodic junctions per cell) relaxed with DFT give an interface formation energy of $0.40\pm0.18$~J\,m$^{-2}$ and a relaxation energy of $-0.073\pm0.012$~eV/atom, larger in magnitude than the $-0.049$~eV/atom of bulk a-\ce{HfO2} under the same protocol, so the junction reorganizes more than the bulk does. Quantities of this kind need an ensemble: the six seeds span $0.20$ to $0.63$~J\,m$^{-2}$, so any single construction would have landed anywhere in that range.

\emph{Doping.} Dopant concentration is a continuous conditioning variable, so a doping series is a sweep rather than a set of separately built models. For the (\ce{Ti}$_x$\ce{Ge}$_{1-x}$)\ce{O2} pseudo-binary that motivates the coating application we generated $11$ compositions from $x=0$ to $1$, ten configurations each, and evaluated all $110$ with DFT (\cref{fig:versatile}(e)). One set of weights tracks the whole series: the absolute energy changes by $2.4$~eV/atom between the endpoints, yet the configurations follow the requested target with a mean absolute error of $30$~meV/atom and a within-composition spread of only $4$--$14$~meV/atom, and they sit close to their own local minima, a subsequent DFT relaxation lowering the energy by $4$--$13$~meV/atom for interior compositions. Energy--volume scans of the relaxed structures give a bulk modulus rising from $30$--$36$ to $\approx76$~GPa as \ce{TiO2} is added, so \ce{TiO2} stiffens the amorphous \ce{GeO2} network. The endpoints bracket what is known: our a-\ce{GeO2} value is $25$--$50\%$ stiffer than the $24.0$~GPa measured for vitreous \ce{GeO2}~\cite{suito1992geo2}, an overstiffness typical of PBE, while the a-\ce{TiO2} endpoint agrees with amorphous \ce{TiO2} films, whose measured $E\approx110$~GPa~\cite{anderson1997tio2} implies $B=E/[3(1-2\nu)]=73$--$92$~GPa for the $\nu=0.25$--$0.30$ typical of oxide glasses~\cite{rouxel2007glass}. No elastic data exist for a-(\ce{Ti},\ce{Ge})\ce{O2}, so the interior of the curve is a prediction. The same conditioning generates $\sim\!10^{4}$-atom coating models of $30\%$ \ce{ZrO2}-doped \ce{Ta2O5} and $27\%$ \ce{SiO2}-doped \ce{HfO2} with the expected first shells---\ce{Ta}--\ce{O} at \SI{1.94}{\angstrom} and \ce{Zr}--\ce{O} at \SI{2.13}{\angstrom}; \ce{Hf}--\ce{O} at \SI{2.09}{\angstrom} with a \ce{Si}--\ce{O} pre-peak at \SI{1.61}{\angstrom}---and in the $9999$-atom cell the two cations keep separate environments in one glass, \ce{Si} four-fold at mean $4.13$ and \ce{Hf} six-fold at mean $6.39$, for first-shell cutoffs of \SI{2.20}{\angstrom} (\ce{Si}--\ce{O}) and \SI{2.90}{\angstrom} (\ce{Hf}--\ce{O}).

A generated configuration is a starting model to hand to DFT or an MLIP: every quantitative number above was obtained after such a relaxation. Generalization is bounded by what the conditioning variables were shown during training, the subject of the next section.

\subsection{Energy conditioning}
\label{sec:conditioning}
\label{sec:coverage}

Generation outside the training set is only useful if it is quantitative: a user requesting a structure $0.05$~eV/atom above the amorphous minimum should receive one. We measure this by generating configurations across a ladder of target excess energies $\Delta E$ above the family minimum, evaluating each with a DFT single point, and regressing achieved on requested excess energy. Perfect calibration gives unit slope; the residual scatter measures the conditional noise of the generator.

How the request is encoded matters more than how large the network is, and this had to be corrected before any of the following numbers could be obtained. Earlier models carried the target in three redundant forms and responded weakly to all of them; supplying the excess energy at generation time, with the weights unchanged, restores the response in every chemistry at once: slopes move from $0.41$ to $1.27$ (\ce{GeO2}), $0.12$ to $1.10$ (\ce{TiO2}), $0.13$ to $1.44$ (\ce{Ti}-\ce{Ge}-\ce{O}), $0.05$ to $1.16$ (\ce{HfO2}) and $-0.04$ to $1.02$ (\ce{SiO2}). The redundancy was then removed rather than worked around, and the models reported here condition on the excess energy alone; the encoding widths that caused it, and why classifier-free guidance is not an alternative, are given in Methods.

\begin{figure}[t]
 \centering
 \includegraphics[width=0.92\linewidth]{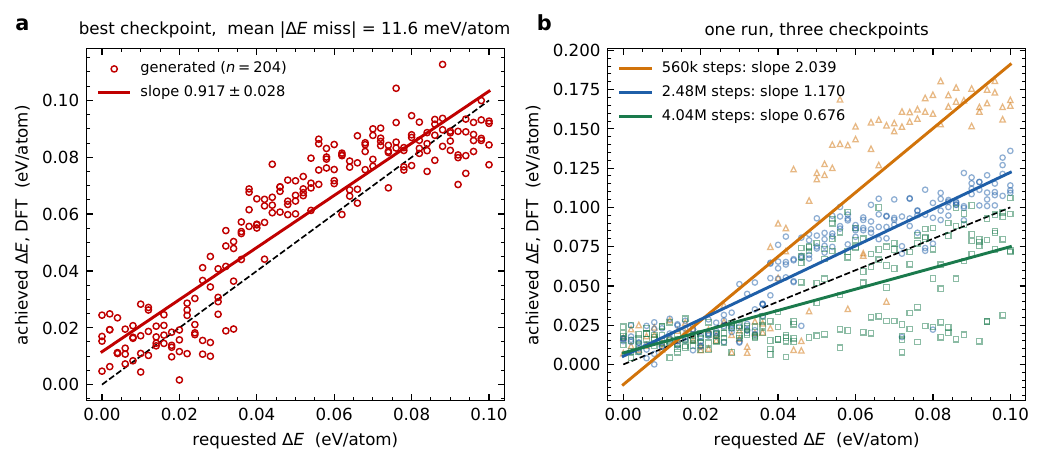}
 \caption{Energy conditioning on amorphous \ce{GeO2}, DFT-verified. Both panels use the dit6 short-range models at a refinement noise scale of \SI{0.02}{\angstrom}, not the production generator of \cref{sec:genverify,sec:explore,sec:versatility}. \textbf{a},~Achieved against requested excess energy for the best of the seven checkpoints scanned, a $606{,}584$-parameter short-range model at $732{,}100$ training steps: $51$ requests from $0$ to $0.100$~eV/atom, four repeats each, \ce{Ge120O240} at \SI{3.650}{\gram\per\cubic\centi\meter}. Dashed line $y=x$, solid line the least-squares fit. \textbf{b},~The same fit for one training run at three checkpoints (\cref{tab:checkpoints}).}
 \label{fig:geo2calib}
\end{figure}

We characterize the resulting accuracy on amorphous \ce{GeO2}, where the measurement can be made dense enough to separate calibration from noise (seven checkpoints, $1{,}020$ DFT single points; Methods). \Cref{fig:geo2calib} shows the result for the best checkpoint. Its slope is $0.917\pm0.028$ and its mean absolute miss over all $204$ of its configurations is $11.6$~meV/atom, with a repeat spread of $7.8$~meV/atom at fixed request.

That repeat spread sets what is achievable. A perfectly calibrated generator with a per-structure spread of $7.8$~meV/atom would still miss $7.8\sqrt{2/\pi}=6.2$~meV/atom from sampling alone, so only $5.4$ of the observed $11.6$ is miscalibration; the DFT contribution is negligible, at $2.8\times10^{-6}$~meV/atom for an energy convergence of $10^{-6}$~eV over $360$ atoms. It also means that models must be ranked by absolute miss and not by slope. Among the checkpoints scanned, the one whose slope is nearest unity, $0.958\pm0.036$, misses $15.2$~meV/atom, worse than the $11.6$ of the $0.917$ model, because its repeat spread is $12.3$ against $7.8$. A perfect slope with loose repeats is not the better conditioner. The sparse-corpus model of \cref{sec:trainset} is the one with the slope nearest unity: trained on one fifth of the data, it reaches a mean absolute miss of $15.2$~meV/atom against the full-corpus model's $11.6$.

The measurement extends to eight chemistries steered by one set of weights: the production generator of \cref{sec:genverify,sec:explore,sec:versatility}, at its pinned refinement noise scale of $10^{-3}$~\AA{} and over requests spanning $0$ to $0.65$~eV/atom, with $13$--$15$ configurations per chemistry and $118$ in total, of which $15$ exceeded the family minimum by more than $2$~eV/atom and were counted as failures and excluded from the fits. These are the production generator's numbers rather than the dense scan's, taken over a wider request range and at the older sampler's pinned refinement noise scale, so the two should be read as separate measurements. Least-squares fits of achieved against requested excess energy give slopes of $0.52$ to $1.54$, correlations of $0.88$ to $0.98$, biases of $0$ to $+0.35$~eV/atom and scatter of $0.025$--$0.066$~eV/atom, where a perfectly calibrated model would give slope $1$ and zero bias. The scatter is small and nearly the same for families whose absolute energies differ by more than $5$~eV/atom. Two systematic offsets remain: \ce{TiO2} and \ce{Si} respond too weakly ($0.59$ and $0.52$) and \ce{HfO2} too strongly ($1.54$), while the other five (\ce{GeO2}, \ce{Ti}-\ce{Ge}-\ce{O}, \ce{SiO2}, \ce{NbO2} and \ce{Zr}-\ce{Ta}-\ce{O}) lie between $0.92$ and $1.06$. A constant slope error can be corrected by rescaling the request, so these are calibration offsets rather than a loss of control, and the high correlations show the response itself is intact.

What limits the response is the range of energies each family's labels span, not the size of the network; coverage is measured here within the conditioning window $\Delta E\le0.8$~eV/atom, beyond which the encoding basis saturates. With the conditioning supplied as an absolute energy, across $540$ DFT single points covering six chemistries only \ce{GeO2} responded---and \ce{GeO2} was the only family whose labels spanned more than $0.5$~eV/atom ($0.577$, with $17\%$ more than $0.3$~eV/atom above its minimum). Every unresponsive family was confined to a narrow range: $0.316$~eV/atom for \ce{TiO2}, and $0.04$--$0.24$~eV/atom per composition for the \ce{Ti}-\ce{Ge}-\ce{O} mixtures. No family had labels between $0.3$ and $0.65$~eV/atom above its minimum, the range a real quench reaches and the one most often requested. A model cannot be steered along a coordinate on which it was shown no variation. Adding data confirmed this in proportion: about $25$ new high-temperature configurations per system, some $1\%$ of the dataset, raised \ce{TiO2} from $0.02$ to $0.12$ and \ce{Ti}-\ce{Ge}-\ce{O} from $0.01$ to $0.13$, while families given no new data did not move. Adding capacity did not. At fixed dataset, conditioning, schedule and step count, a $14.2$M-parameter version of the $600$k backbone---a $70$-function radial basis at \SI{0.10}{\angstrom} resolution against \SI{0.44}{\angstrom}, a wider hidden space including $\ell{=}2$ channels, and a fourth convolution layer---reached a validation error of $0.10227$ against $0.10427$ at the same $79{,}256$ steps: $2\%$ better at $16.4\times$ the wall time per step, with no gain in conditioning. A $23\times$ increase in parameters bought $2\%$ in training error.

Outright failure follows the same rule: it happens when the request points into a region of the label distribution the training set left empty. \ce{Nb} is $0.52\%$ of the cations in the corpus, and a-\ce{NbO2} is the only system we have tried that cannot be generated---its atoms collapse to the sampler's \SI{1.1}{\angstrom} steric floor at every size we tried, while \ce{GeO2}, \ce{TiO2}, \ce{Ti}-\ce{Ge}-\ce{O} and \ce{HfO2} are physical in every case. Capacity makes it worse rather than better: across three model sizes the collapse rate runs from $0/8$ for a half-width model to $8/8$ for the largest, because a larger model extrapolates more confidently into a composition it was barely shown, and training it twice as long changed nothing. The same pattern sets a ceiling on the requested energy within a single chemistry. In a census of $2{,}211$ \ce{Ge120O240} configurations over eleven checkpoints and requests from $0$ to $0.5$~eV/atom, collapse concentrates at the high-$\Delta E$ end, where the labels are sparsest, and four checkpoints---among them both of the sparse-corpus model, whose trajectories are the most widely spread---produce none anywhere in the range. Conditioning fidelity and structural validity have therefore to be reported together: a requested energy is not usable merely because it lies inside the trained range, and every generated configuration should be screened on minimum interatomic distance before it is used.

\section{Discussion}
\label{sec:discussion}

We have presented DiT, an equivariant denoising-diffusion model for amorphous oxides that learns the configurational ensemble of a family rather than any single structure. It generates configurations whose partial radial distribution functions and coordination statistics match DFT references from a few hundred atoms to $3\times10^{5}$, in wall-clock times within a small factor of empirical pair potentials and far below DFT-MD at the same size. For a material defined by an ensemble rather than by one structure, the trained model is itself the database, and a faster and more flexible one than a tabulated list.

Learning that ensemble does not require exhaustive sampling of it. A model trained on $1{,}781$ configurations from $430$ DFT trajectories---one fifth of the comparison set, with no family contributing more than $49$ trajectories---reproduces the same structural observables and steers to a requested energy with a mean absolute error of $15.2$~meV/atom against $11.6$ for the full-corpus model. (Its calibration slope is the nearest unity of any checkpoint scanned, but by the criterion of \cref{sec:conditioning} that is not a merit: its repeat spread is larger, which is why its absolute miss is worse.) Over $201$ requested energies spanning $0$ to $0.5$~eV/atom it produced no collapsed configuration at all, where the larger-corpus production generator collapsed $59$ of $201$. What matters is that the trajectories cover the ensemble, not that there are many of them: a few hundred well-spread DFT trajectories are enough, and the effort is better spent spreading them than lengthening them.

Measured against a fixed reference rather than against the lowest energy in its own training set, generation becomes discovery. The model located an amorphous \ce{Zr}-\ce{Ta}-\ce{O} structure lying $36$~meV/atom below any previously reported, about a third of that family's amorphization energy of $+110$~meV/atom, so the structure previously taken as the minimum lay measurably above it. Conditioned generation followed by a single DFT relaxation is therefore a way of searching the amorphous energy landscape~\cite{stillinger1995topographic,debenedetti2001supercooled}, at a cost set by the relaxations rather than by the model, with the division of labour that the model proposes configurations and DFT decides.

The learned ensemble also extends past its training conditions, to non-stoichiometric compositions, requested mass densities, oxide--oxide interfaces and doped compositions. For the applications motivating this work the useful output is a complete periodic atomic model at a size and composition that could not otherwise be built: a coherent \ce{SiO2}/(\ce{Ti},\ce{Ge})\ce{O2} multilayer at $138\,345$ atoms, six independent a-\ce{SiO2}/a-\ce{HfO2} interfaces from which a formation energy can be averaged rather than asserted, and a continuous doping series whose predicted stiffening from $30$--$36$ to $76$~GPa across the (\ce{Ti}$_x$\ce{Ge}$_{1-x}$)\ce{O2} pseudo-binary has no experimental counterpart yet. These are starting models for subsequent first-principles or MLIP calculations, which is how every quantitative number here was obtained.

Two methodological results apply beyond this model. First, the training objective does not rank generative quality: our lowest-error model generated no usable structure in $100$ attempts, and a single run lowers its loss monotonically while its conditioning accuracy peaks and then decays. The objective is dominated by a noise range in which our architectures agree and is nearly blind to the narrow range that fixes bond lengths, so the informative monitor is the loss resolved by noise scale, not its mean. Second, conditioning accuracy follows the range of energies the training labels span and is almost unchanged by a $23\times$ increase in parameters, so effort is better spent on what is sampled than on network size; compute-optimal scaling studies and information-theoretic analyses of atomistic datasets reach the same conclusion~\cite{kaplan2020scaling,hoffmann2022training,schwalbekoda2025information}. When a chemistry fails at generation, check its coverage first: ours failed on the scarcest cation, and added capacity made it worse.

Three directions follow. The first is the long-range branch, which reaches the lowest training error of any variant we built yet improves nothing about generation; where it enters the network decides whether the model is usable at all. The second is the training set, which we are extending where it is thinnest---at energies well above each composition's minimum---rather than enlarging the network. The third is cost: a two-stage arrangement in which a small model produces the configuration and a larger one refines the last denoising steps, as in image generation~\cite{dosovitskiy2021vit,peebles2023dit,esser2024sd3,yu2025pixeldit,li2025backtobasics}, together with fine-tuning against a reward on relaxed properties. One further question is where the amorphous phase ends: in \ce{GeO2} the glass and $\alpha$-quartz are degenerate to $0.8$~meV/atom, so the crystallization boundary cannot be located from energy at all, and deciding when an annealed film has begun to crystallize needs structural order parameters. That is what would turn the annealing window of \cref{sec:intro} into a design variable.


\section{Methods}
\label{sec:methods}

\subsection{Diffusion-process setup}
\label{sec:setup}
\label{sec:vs}

A diffusion model predicts the noise that must be subtracted from the current configuration to make it slightly more probable under the target distribution. Generation starts from random atomic positions and iterates that denoising step until the configuration is consistent with the distribution. Because every training example pairs a clean configuration with a noisy one, an arbitrary starting configuration is simply a point on a noise scale the model was trained across, and successive steps move it along the gradient of the learned log-density onto the training distribution. The cost is the few thousand steps needed to traverse that scale; what it buys, recovery from an arbitrary initial configuration, is what an amorphous-structure generator requires.

Given an atomic configuration of $N$ atoms with positions $\bm{x} \in \mathbb{R}^{N \times 3}$ in a periodic cell, the diffusion forward process adds isotropic Gaussian noise
\begin{equation}
 \tilde{\bm{x}} = \bm{x} + \sigma \, \bm{\varepsilon}, \qquad \bm{\varepsilon} \sim \mathcal{N}(0,\bm{I}),
\end{equation}
with a per-structure noise scale $\sigma$ sampled uniformly from $[\sigma_{\min}, \sigma_{\max}]$ during training ($\sigma_{\min}{=}10^{-3}$, $\sigma_{\max}{=}0.75$~\AA{} by default). An equivariant neural network $\hat{\bm{\varepsilon}}_{\bm{\theta}}(\tilde{\bm{x}}, \sigma, \bm{c})$, parameterized by weights $\bm{\theta}$ and decorated with a caret to mark it as a prediction, is trained to estimate the noise displacement $\sigma\bm{\varepsilon}$ from the perturbed positions, the noise scale, and a vector of scalar conditioning variables $\bm{c}$ (potential energy per atom, mass density, etc.) by minimizing the mean squared error
\begin{equation}
 \mathcal{L}(\bm{\theta}) = \mathbb{E}_{\bm{x},\sigma,\bm{\varepsilon}} \bigl\lVert \hat{\bm{\varepsilon}}_{\bm{\theta}}(\tilde{\bm{x}}, \sigma, \bm{c}) - \sigma\bm{\varepsilon} \bigr\rVert^{2}.
 \label{eq:loss}
\end{equation}
This is the standard DDPM~\cite{ho2020denoising} $\mathcal{L}_{\mathrm{simple}}$ objective in its displacement-prediction parameterization; up to an overall $\sigma^{2}$ factor it is equivalent to the unit-variance $\bm{\varepsilon}$-prediction form used by EDM~\cite{edm}.
\paragraph*{Normalization of the reported loss.} Every loss quoted in this paper is a mean over atoms and over the three Cartesian components, in \AA{}$^{2}$, with $\sigma$ drawn uniformly from $[10^{-3},0.75]$~\AA{}. Two objectives are in circulation in this work and they differ by roughly a factor of three, so their values must never be compared: the \emph{unweighted} form of \cref{eq:loss}, and a $1/\sigma^{2}$-weighted variant that multiplies the per-element error by $\sigma^{-2}$ to match the $\bm{\varepsilon}$-prediction convention and so removes the $\sigma\!\approx\!\sigma_{\max}$ domination of the gradient. Unweighted values fall near $0.08$--$0.09$; $1/\sigma^{2}$-weighted values near $0.27$--$0.33$. Each number below states which objective it is on, and comparisons are made only within one objective, one dataset and one batch size.
At inference time, generation starts from a random initial configuration and iteratively subtracts the predicted noise according to a decreasing $\sigma$ schedule (the EDM (Karras) $\sigma$ schedule~\cite{edm}), optionally followed by a brief refinement phase held at the floor $\sigma_{\min}$ with no further noise added, which cleans up residual structure. Improvements to the basic DDPM sampler introduced by Nichol \& Dhariwal~\cite{nichol2021improved} and the conditioning techniques of Choi \emph{et~al.}~\cite{choi2021ilvr} inform the implementation, though the present model uses the simpler EDM schedule throughout.

There is also a physical correspondence between the diffusion process and the formation of an amorphous structure, and it is the reason this class of model suits the problem. An amorphous configuration of a covalent oxide is, to first approximation, a \emph{thermally distorted} version of one of its crystalline polymorphs~\cite{mavracic2018tio2,harrelson2021niobium}: the same coordination polyhedra and the same local bond-network topology, with random angular and bond-length perturbations imposed by the quench history, a sparse set of point defects (oxygen vacancies, interstitials, miscoordinated atoms), and broken periodicity at the medium-range scale. The forward diffusion process of DDPM~\cite{ho2020denoising} adds isotropic Gaussian noise to atomic positions in exactly the way thermal motion does: the variance $\sigma^{2}$ plays the role of $k_{\text{B}}T$ divided by an effective spring constant, and the reverse process learns to remove that noise. Training jointly on crystalline and amorphous DFT configurations therefore exposes the network to the full spectrum of local environments and their relative energetic ordering, and teaches it a single denoiser that maps high-noise, high-effective-temperature inputs back toward low-noise, low-disorder basins along a continuous schedule~\cite{liu2023generative}. Sampling at intermediate $\sigma$ gives intermediate distortion, corresponding to a configuration at a finite effective temperature, which is the regime in which amorphous oxides are prepared and annealed. The training objective is in this sense a mathematical analogue of the thermal process that produces the amorphous distribution. The analogy motivates the method but should not be pushed quantitatively, and it fails in two identifiable ways. Forward diffusion adds independent isotropic noise to every atom, whereas thermal displacements are correlated through the dynamical matrix, so the correspondence is exact only for a single Einstein-like mode and not for a real phonon spectrum. And a glass differs from its crystalline polymorph in medium-range topology, not only in bond-angle and bond-length scatter: the edge-sharing polyhedra and three-coordinated oxygens reported above are not thermal distortions of any polymorph motif. The noise process supplies the local disorder; the medium-range topology has to be learned from the data.

\subsection{Equivariant network architecture}
\label{sec:arch}

\begin{figure}[t]
 \centering
 \includegraphics[width=0.99\linewidth]{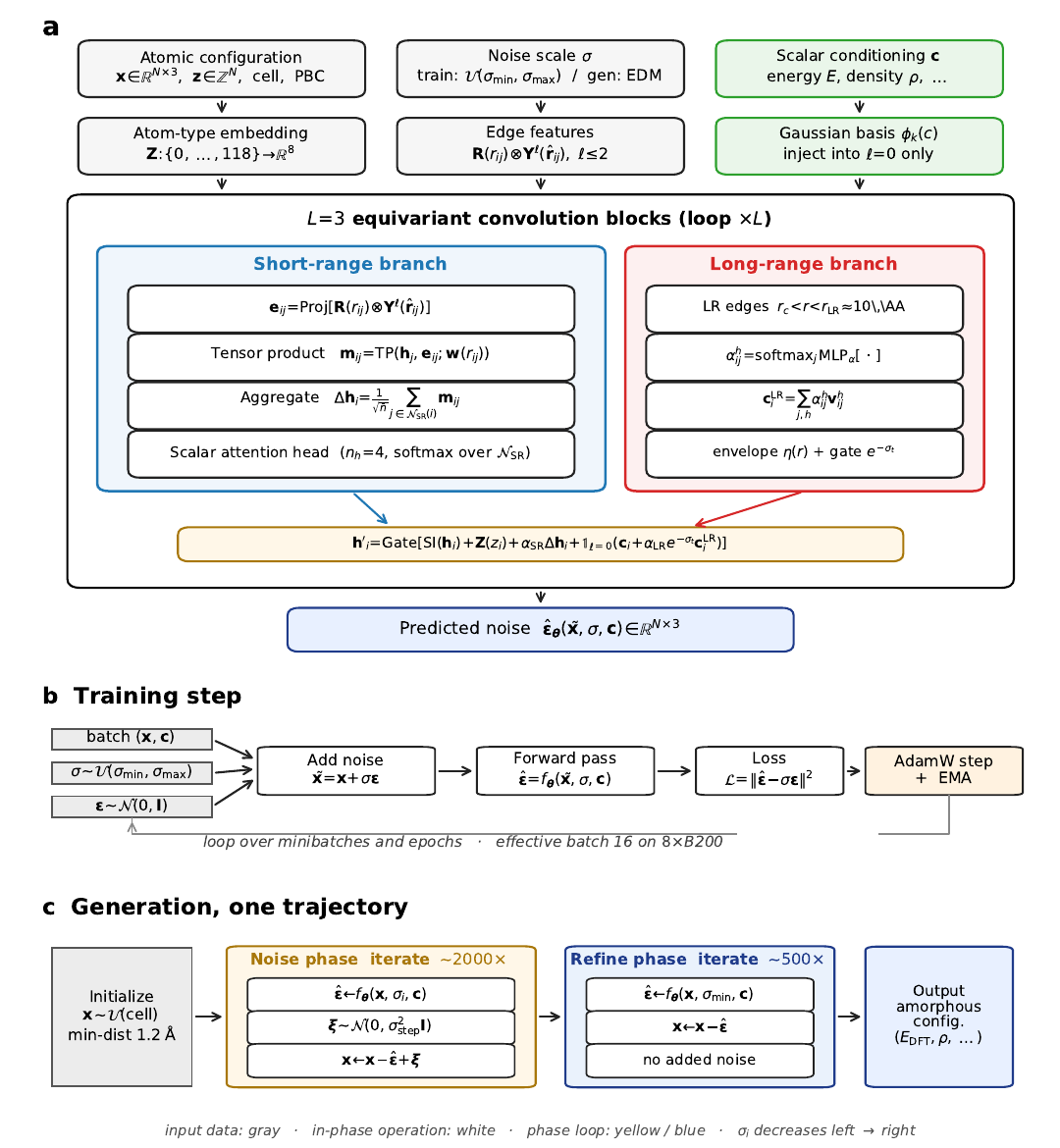}
 \caption{Architecture and end-to-end pipeline. \textbf{a},~Forward pass: the configuration, noise scale $\sigma$ and scalar conditioning $\mathbf{c}$ feed an embedding row and $L\!=\!3$ equivariant convolution blocks, each with a short-range branch (blue) and, for the dual-cutoff variant only, a long-range branch (red). \textbf{b},~Training step; the dashed loop returns to the next minibatch. \textbf{c},~One generation trajectory: a noise phase (yellow) with a Brownian kick per step, then a refine phase (blue) at the floor $\sigma_{\min}$ with no added noise. Defaults and update equations are given in the text.}
 \label{fig:arch}
\end{figure}

The network $\hat{\bm{\varepsilon}}_{\bm{\theta}}$ is an $SE(3)$-equivariant message-passing graph network built on the \textsc{e3nn} library~\cite{e3nn}. The models reported here condition on the excess energy alone, encoded at a fixed width (\cref{sec:conditioning}), although the conditioning mechanism itself accepts further scalars. Generation is size independent because the network predicts a per-atom displacement, so one set of weights generates cells with different atom numbers. \Cref{fig:arch} sketches the overall architecture. Atoms are nodes; edges are introduced between every pair within a cutoff radius $r_c$ (typically \SIrange{5}{7}{\angstrom}), and a second long-range edge set within $r_{\text{LR}}$ (typically \SI{10}{\angstrom}) is built for the dual-cutoff variant. Each node carries a direct sum of irreducible-representation features $\bm{h}_i \in \bigoplus_\ell \mathbb{R}^{m_\ell (2\ell+1)}$, with default multiplicities $m_0{=}64$, $m_1{=}32$, retaining both scalar ($\ell{=}0$) and vector ($\ell{=}1$) channels. Edges carry Bessel-basis radial embeddings $\bm{R}(r_{ij})$ together with component-normalized spherical-harmonic angular embeddings $\bm{Y}^\ell(\hat{\bm{r}}_{ij})$, projected through a small linear layer into the irreps of $\bm{e}_{ij}$ (default $4{\times}\ell{=}0 + 4{\times}\ell{=}1 + 2{\times}\ell{=}2$). Parity is not tracked: every irreducible representation carries even parity, including the odd-$\ell$ channels. The network is therefore equivariant to the proper rotations and translations of $SE(3)$, and to permutations of like atoms, but not to inversion or mirror operations, whose invariance it must instead learn from the data. We verified this in double precision: the relative error $\lVert\hat{\bm{\varepsilon}}(\bm{Q}\tilde{\bm{x}})-\bm{Q}\hat{\bm{\varepsilon}}(\tilde{\bm{x}})\rVert/\lVert\hat{\bm{\varepsilon}}(\tilde{\bm{x}})\rVert$ is $5\times10^{-11}$ for proper rotations ($\det\bm{Q}{=}{+}1$), exactly zero for translations, and $1.0\times10^{-2}$ for inversion and for mirrors ($\det\bm{Q}{=}{-}1$). Relabelling the odd-$\ell$ irreps with odd parity removes the improper error identically, and the parity selection rules then forbid $3.3\times10^{4}$ of the $5.9\times10^{5}$ parameters. The observables in this work, such as partial radial distribution functions, coordination numbers and distributions, bond angles, and DFT energies, are invariant under inversion and mirrors. A chirality-resolved order parameter, or any other quantity that distinguishes a structure from its mirror image, would require the improper operations to be included.

The backbone used for every configuration reported in this paper is the \texttt{graphite}\cite{graphite} baseline, which reproduces the original NequIP convolution faithfully and is described below. We also built a \texttt{dual\_cutoff} backbone that augments it with a long-range branch over a second, wider edge set within $r_{\text{LR}}$; it shares the outer message-passing scaffold and differs in how the per-edge message is parameterized and how distant pairs enter the node update. As shown in \cref{sec:lossquality}, its variants reach the lowest training error we have measured and do not improve generation quality, so no result uses it in this work. The present implementation builds on, and is verified against, two open-source codebases that we acknowledge throughout this work: the LLNL \textsc{graphite} graph-neural-network library~\cite{graphite}, whose Interaction layer the \texttt{graphite} backbone reproduces line-for-line, and the \textsc{DM2} amorphous-oxide diffusion-model implementation~\cite{dm2}, built on the score-based denoising method of Hsu \emph{et~al.}~\cite{hsu2024score}, whose training pipeline and noise-rattle transforms seeded the present codebase.

\subsubsection{Generic NequIP message-passing}

The baseline equivariant update follows the NequIP~\cite{nequip} construction. For a destination node $i$ with neighbour set $\mathcal{N}(i)$ and per-edge radial-MLP weights $\bm{w}(r_{ij})$, the edge feature, message, and aggregation are
\begin{align}
 \bm{e}_{ij} &= \mathrm{Proj}\!\big[\bm{R}(r_{ij}) \otimes \bm{Y}^{\ell}(\hat{\bm{r}}_{ij})\big], \label{eq:edge}\\ \bm{m}_{ij} &= \mathrm{TP}\!\big(\bm{h}_j,\, \bm{e}_{ij};\, \bm{w}(r_{ij})\big), \label{eq:msg}\\ \Delta \bm{h}_i &= \frac{1}{\sqrt{\bar{n}}}\sum_{j\in\mathcal{N}(i)} \bm{m}_{ij}, \label{eq:agg}
\end{align}
where $\mathrm{TP}$ is the equivariant fully-connected tensor product of \textsc{e3nn}~\cite{e3nn} with per-edge weights from a small radial MLP, and the $1/\sqrt{\bar{n}}$ factor normalizes by the average neighbour count $\bar{n}$. The per-layer node update combines the convolutional contribution with two residual paths and the scalar conditioning vector $\bm{c}_i$,
\begin{equation}
 \bm{h}'_i = \mathrm{Gate}\!\left[\mathrm{SI}(\bm{h}_i) + \mathrm{Z}(z_i) + \alpha_{\mathrm{SR}}\,\Delta\bm{h}_i + \mathbb{1}_{\ell=0}\,\bm{c}_i\right], \label{eq:update}
\end{equation}
where $\mathrm{SI}$ is an equivariant self-interaction linear, $\mathrm{Z}(z_i)$ projects the static atom-type embedding, $\mathrm{Gate}$ is an \textsc{e3nn} equivariant gated nonlinearity~\cite{e3nn}, and $\mathbb{1}_{\ell=0}\bm{c}_i$ injects the scalar conditioning into the $\ell{=}0$ multiplet only, preserving the equivariance of higher-order channels. Each layer carries a learnable SkipInit scalar $\alpha_{\mathrm{SR}}\in\mathbb{R}$ initialized to zero, so at the start of training only the residual stream is active and the convolutional contribution ramps in as the model learns.

\subsubsection{Graphite backbone}
\label{sec:arch:graphite}

The \texttt{graphite} backbone is the original NequIP convolution as distributed in the \textsc{graphite} library~\cite{graphite}. The pre- and post-tensor-product projections are dense linear maps that take the full hidden multiplicity in and out of the tensor product without any bottleneck, faithfully reproducing the construction of Batzner~\emph{et~al.}~\cite{nequip}. At the default irreps $64{\times}0e + 32{\times}1e$ the resulting model has $606{,}584$ trainable parameters at the settings used throughout this paper. We retain this backbone as a calibration reference against which the dual-cutoff variant is measured.

\subsection{Multi-property conditioning}

Scalar conditioning variables (energy, mass density, cooling rate, pressure, $\ldots$) are projected through Gaussian radial bases $\phi_k(c)$ centered uniformly across the property range, then through a per-property MLP into the node-feature dimension. Encoding width is not a free choice. Models that carried the target in three redundant forms---an absolute energy, an excess $\Delta E$ above the composition minimum, and an effective temperature $T_{\mathrm{eff}}=\Delta E/\tfrac{3}{2}k_{\mathrm{B}}$ obtained from it by equipartition---encoded them at different widths, $0.026$~eV/atom for the excess against $0.484$~eV/atom for the absolute energy. Because the three were dropped independently during training, most samples carried the narrower encoding, so the model learned the excess and its handling of absolute energy degraded. In the models reported here the absolute-energy and effective-temperature branches are absent, and $\Delta E$ is encoded in $32$ Gaussians of width $0.032$~eV/atom spanning $0$ to $1$~eV/atom. Amplifying a weak response at sampling time is not an alternative: classifier-free guidance~\cite{ho2022classifierfree} applied to the unchanged model over $300$ DFT-evaluated configurations sweeps the slope from $0.57$ through $1.86$ to $3.46$, passing through unity near a guidance strength of $1.7$, but the scatter stays near $0.25$~eV/atom, about five times that of a model trained with the conditioning always present, and failures appear at the highest setting. Per the design choice we labelled ``Fix~1'' during development, the conditioning is added \emph{only} into the $\ell{=}0$ (scalar) channels (preserving the equivariance of higher-order channels) and zero-padded into the remaining irreducible representations.

\subsection{Generation loop}

Generation follows panel~\textbf{c} of \cref{fig:arch}: $\sigma$ is swept along the EDM power-law schedule of Karras~\emph{et~al.}~\cite{edm} with $\rho{=}7$, $\sigma_{\max}{=}0.75$~\AA{}, $\sigma_{\min}{=}10^{-3}$~\AA{}, over $\sim$2000 noise steps and $\sim$500 refine steps. The user supplies target values for the conditioning variables (e.g., a target $E_{\text{DFT}}$ in eV/atom and a target mass density in g/cm$^{3}$), biasing the trajectory toward configurations consistent with those targets. Across the schedule the configuration progresses from a structureless random cloud through partially bonded states with emerging coordination polyhedra and short-range order, settling onto the manifold of physically reasonable amorphous configurations during the noiseless refine phase.


\subsection{Reference energies}
Both reference energies were obtained by relaxing $216$--$384$-atom supercells of every relevant polymorph under the same settings as the amorphous calculations, so that the comparison of \cref{sec:explore} is minimum to minimum under one protocol. The previous \ce{Zr}-\ce{Ta}-\ce{O} record was re-relaxed under that identical protocol and recovered to within $1$~meV/atom.

\subsection{Conditioning calibration protocol}
For each of seven checkpoints we generated \ce{Ge120O240} at \SI{3.650}{\gram\per\cubic\centi\meter} at $51$ requested excess energies from $0$ to $0.100$~eV/atom in $0.002$ steps, with two to four repeats per request, and evaluated every configuration with a DFT single point. Of $1{,}020$ calculations $1{,}019$ converged, and $1{,}016$ configurations passed the \SI{1.45}{\angstrom} chemistry screen.

\subsection{Software implementation}
\label{sec:soft}

The complete pipeline (training, generation, and analysis) is implemented in PyTorch on top of \texttt{torch\_geometric}, \texttt{torch\_scatter} and \textsc{e3nn}, and its generation path, that is the network definition, the checkpoint loader, the sampler, the structure and neighbour-list builders and the radial-distribution and coordination analysis, is deposited at \url{https://github.com/jiangjun413/dit-amorphous} (tag \texttt{v1.1}), which is what is required to load the deposited checkpoints and generate with them. Training supports single-GPU and multi-GPU distributed data-parallel execution, with on-device loss accumulation to avoid the per-step host--device synchronization that otherwise dominates wall time on fast accelerators, and a CPU-portable variant of every checkpoint is saved alongside the GPU version. Beyond the command-line training and generation interfaces, the full package includes a Streamlit-based web user interface intended for non-expert users: a sidebar lets the user load a trained model, specify a target composition (atom counts per element), pick a mass density and target energy, and click ``generate''. Live intermediate configurations and the final RDF / coordination analysis are displayed in the main panel, and the final structure can be downloaded as a VASP POSCAR.


\section*{Data availability}
The data supporting this work are deposited at \url{https://github.com/jiangjun413/dit-amorphous}, tagged \texttt{v1.1}, each file listed with its SHA-256 checksum: the trained DiT model checkpoints, the diffusion-generated amorphous oxide configurations reported in \cref{sec:rdf,sec:assistedopt,sec:versatility}, the record amorphous structure of \cref{sec:explore}, the per-configuration DFT energies underlying the calibration of \cref{sec:conditioning}, the reference-energy table used to define the two references of \cref{sec:assistedopt}, the training and validation loss curves of the three compared backbones, and the structural-verification output behind \cref{tab:liteverify}. The DFT training and validation trajectories, and the raw electronic-structure output from which they were derived, exceed what a code repository can hold and are available from the corresponding author on reasonable request.

\section*{Code availability}
The generation code, that is the network definition, the checkpoint loader, the sampler and the structural analysis, is deposited under the MIT licence with the data at the repository cited above (tag \texttt{v1.1}), and is what is needed to load the deposited checkpoints and reproduce the generation of the configurations reported here. The training pipeline, the dataset construction code, the web user interface and the configuration file of every reported run are available from the corresponding author on reasonable request.
 
\section*{Author contributions}
J.J. conceived the study, designed and implemented the DiT architecture and training pipeline, performed all density-functional theory reference calculations and diffusion-model training runs, analysed the data, prepared all figures, and drafted the manuscript; I.B. contributed DFT calculations for validation; J.N.F. worked on mathematics underlying diffusion models and editing the manuscript, and H.-P.C. conceived and directed the project and validated the model.

\section*{Competing interests}
The authors declare no conflict of interest.

\section*{Acknowledgments}
This work is supported by the National Science Foundation (NSF) under Award Nos.\ PHY-2426460, PHY-2430436 and Moore Foundation FOUND-PT/Stanford 63839614-351582. The Computations were performed using the resources of the National Energy Research Scientific Computing Center (NERSC) and the University of Florida Research Computing HiPerGator cluster.

\bibliographystyle{unsrtnat}
\bibliography{ref}

@inproceedings{joshi2025allatom,
  author       = {Joshi, Chaitanya K. and Fu, Xiang and Liao, Yi-Lun
                  and Gharakhanyan, Vahe and Miller, Benjamin Kurt and
                  Sriram, Anuroop and Ulissi, Zachary W.},
  title        = {All-atom Diffusion Transformers: Unified Generative
                  Modelling of Molecules and Materials},
  booktitle    = {Proceedings of the 42nd International Conference on
                  Machine Learning (ICML)},
  series       = {PMLR},
  volume       = {267},
  pages        = {28393--28417},
  year         = {2025},
  publisher    = {PMLR},
  url          = {https://proceedings.mlr.press/v267/joshi25a.html},
}

@misc{li2025backtobasics,
  author       = {Li, Tianhong and He, Kaiming},
  title        = {Back to Basics: Let Denoising Generative Models
                  Denoise},
  year         = {2025},
  month        = nov,
  eprint       = {2511.13720},
  archivePrefix= {arXiv},
  primaryClass = {cs.CV},
  url          = {https://arxiv.org/abs/2511.13720},
  note         = {Preprint, 17~Nov.~2025},
}

@misc{yu2025pixeldit,
  author       = {Yu, Yongsheng and Xiong, Wei and Nie, Weili and
                  Sheng, Yichen and Liu, Shiqiu and Luo, Jiebo},
  title        = {{PixelDiT}: Pixel Diffusion Transformers for Image
                  Generation},
  year         = {2025},
  month        = nov,
  eprint       = {2511.20645},
  archivePrefix= {arXiv},
  primaryClass = {cs.CV},
  url          = {https://arxiv.org/abs/2511.20645},
  note         = {Preprint, 25~Nov.~2025},
}

@article{nequip,
  author       = {Batzner, Simon and Musaelian, Albert and Sun, Lixin
                  and Geiger, Mario and Mailoa, Jonathan P. and
                  Kornbluth, Mordechai and Molinari, Nicola and
                  Smidt, Tess E. and Kozinsky, Boris},
  title        = {{E(3)}-equivariant graph neural networks for
                  data-efficient and accurate interatomic potentials},
  journal      = {Nature Communications},
  volume       = {13},
  number       = {1},
  pages        = {2453},
  year         = {2022},
  doi          = {10.1038/s41467-022-29939-5},
  url          = {https://doi.org/10.1038/s41467-022-29939-5},
}

@inproceedings{mace,
  author       = {Batatia, Ilyes and Kov{\'a}cs, D{\'a}vid P. and
                  Simm, Gregor N. C. and Ortner, Christoph and
                  Cs{\'a}nyi, G{\'a}bor},
  title        = {{MACE}: Higher Order Equivariant Message Passing
                  Neural Networks for Fast and Accurate Force Fields},
  booktitle    = {Advances in Neural Information Processing Systems
                  (NeurIPS)},
  year         = {2022},
  eprint       = {2206.07697},
  archivePrefix= {arXiv},
  url          = {https://arxiv.org/abs/2206.07697},
}

@inproceedings{edm,
  author       = {Karras, Tero and Aittala, Miika and Aila, Timo and
                  Laine, Samuli},
  title        = {Elucidating the Design Space of Diffusion-Based
                  Generative Models},
  booktitle    = {Advances in Neural Information Processing Systems
                  (NeurIPS)},
  year         = {2022},
  eprint       = {2206.00364},
  archivePrefix= {arXiv},
  url          = {https://arxiv.org/abs/2206.00364},
}

@inproceedings{sohl2015deep,
  author       = {Sohl-Dickstein, Jascha and Weiss, Eric and
                  Maheswaranathan, Niru and Ganguli, Surya},
  title        = {Deep Unsupervised Learning using Nonequilibrium
                  Thermodynamics},
  booktitle    = {International Conference on Machine Learning
                  (ICML)},
  pages        = {2256--2265},
  year         = {2015},
  organization = {PMLR},
  eprint       = {1503.03585},
  archivePrefix= {arXiv},
  url          = {https://arxiv.org/abs/1503.03585},
}

@article{ho2020denoising,
  author       = {Ho, Jonathan and Jain, Ajay and Abbeel, Pieter},
  title        = {Denoising Diffusion Probabilistic Models},
  journal      = {Advances in Neural Information Processing Systems
                  (NeurIPS)},
  volume       = {33},
  pages        = {6840--6851},
  year         = {2020},
  eprint       = {2006.11239},
  archivePrefix= {arXiv},
  url          = {https://arxiv.org/abs/2006.11239},
}

@inproceedings{nichol2021improved,
  author       = {Nichol, Alexander Quinn and Dhariwal, Prafulla},
  title        = {Improved Denoising Diffusion Probabilistic Models},
  booktitle    = {International Conference on Machine Learning
                  (ICML)},
  pages        = {8162--8171},
  year         = {2021},
  organization = {PMLR},
  eprint       = {2102.09672},
  archivePrefix= {arXiv},
  url          = {https://arxiv.org/abs/2102.09672},
}

@inproceedings{song2021scoresde,
  author       = {Song, Yang and Sohl-Dickstein, Jascha and Kingma,
                  Diederik P. and Kumar, Abhishek and Ermon, Stefano
                  and Poole, Ben},
  title        = {Score-Based Generative Modeling Through Stochastic
                  Differential Equations},
  booktitle    = {International Conference on Learning Representations
                  (ICLR)},
  year         = {2021},
  eprint       = {2011.13456},
  archivePrefix= {arXiv},
  url          = {https://arxiv.org/abs/2011.13456},
}

@inproceedings{choi2021ilvr,
  author       = {Choi, Jooyoung and Kim, Sungwon and Jeong,
                  Yonghyun and Gwon, Youngjune and Yoon, Sungroh},
  title        = {{ILVR}: Conditioning Method for Denoising Diffusion
                  Probabilistic Models},
  booktitle    = {Proceedings of the IEEE/CVF International Conference
                  on Computer Vision (ICCV)},
  pages        = {14367--14376},
  year         = {2021},
  eprint       = {2108.02938},
  archivePrefix= {arXiv},
  url          = {https://arxiv.org/abs/2108.02938},
}

@article{dhariwal2021diffusion,
  author       = {Dhariwal, Prafulla and Nichol, Alexander},
  title        = {Diffusion Models Beat {GANs} on Image Synthesis},
  journal      = {Advances in Neural Information Processing Systems
                  (NeurIPS)},
  volume       = {34},
  pages        = {8780--8794},
  year         = {2021},
  eprint       = {2105.05233},
  archivePrefix= {arXiv},
  url          = {https://arxiv.org/abs/2105.05233},
}

@inproceedings{liu2023generative,
  author       = {Liu, Chengyi and Fan, Wenqi and Liu, Yunqing and
                  Li, Jiatong and Li, Hang and Liu, Hui and Tang,
                  Jiliang and Li, Qing},
  title        = {Generative Diffusion Models on Graphs: Methods and
                  Applications},
  booktitle    = {Proceedings of the 32nd International Joint
                  Conference on Artificial Intelligence (IJCAI)},
  pages        = {6702--6711},
  year         = {2023},
  doi          = {10.24963/ijcai.2023/751},
  url          = {https://doi.org/10.24963/ijcai.2023/751},
}

@misc{e3nn,
  author       = {Geiger, Mario and Smidt, Tess},
  title        = {{e3nn}: {Euclidean} Neural Networks},
  year         = {2022},
  eprint       = {2207.09453},
  archivePrefix= {arXiv},
  primaryClass = {cs.LG},
  url          = {https://arxiv.org/abs/2207.09453},
}

@article{allegro,
  author       = {Musaelian, Albert and Batzner, Simon and Johansson,
                  Anders and Sun, Lixin and Owen, Cameron J. and
                  Kornbluth, Mordechai and Kozinsky, Boris},
  title        = {Learning local equivariant representations for
                  large-scale atomistic dynamics},
  journal      = {Nature Communications},
  volume       = {14},
  number       = {1},
  pages        = {579},
  year         = {2023},
  doi          = {10.1038/s41467-023-36329-y},
  url          = {https://doi.org/10.1038/s41467-023-36329-y},
}

@inproceedings{dosovitskiy2021vit,
  author       = {Dosovitskiy, Alexey and Beyer, Lucas and Kolesnikov,
                  Alexander and Weissenborn, Dirk and Zhai, Xiaohua and
                  Unterthiner, Thomas and Dehghani, Mostafa and Minderer,
                  Matthias and Heigold, Georg and Gelly, Sylvain and
                  Uszkoreit, Jakob and Houlsby, Neil},
  title        = {An Image is Worth 16x16 Words: Transformers for Image
                  Recognition at Scale},
  booktitle    = {International Conference on Learning Representations
                  (ICLR)},
  year         = {2021},
  eprint       = {2010.11929},
  archivePrefix= {arXiv},
  url          = {https://arxiv.org/abs/2010.11929},
}

@inproceedings{xie2022cdvae,
  author       = {Xie, Tian and Fu, Xiang and Ganea, Octavian-Eugen and
                  Barzilay, Regina and Jaakkola, Tommi},
  title        = {Crystal Diffusion Variational Autoencoder for Periodic
                  Material Generation},
  booktitle    = {International Conference on Learning Representations
                  (ICLR)},
  year         = {2022},
  eprint       = {2110.06197},
  archivePrefix= {arXiv},
  url          = {https://arxiv.org/abs/2110.06197},
}

@article{zeni2025mattergen,
  author       = {Zeni, Claudio and Pinsler, Robert and Z{\"u}gner,
                  Daniel and Fowler, Andrew and Horton, Matthew and Fu,
                  Xiang and Wang, Zilong and Shysheya, Aliaksandra and
                  Crabb{\'e}, Jonathan and Ueda, Shoko and Sordillo,
                  Roberto and Sun, Lixin and Smith, Jake and Nguyen,
                  Bichlien and Schulz, Hannes and Lewis, Sarah and Huang,
                  Chin-Wei and Lu, Ziheng and Zhou, Yichi and Yang, Han
                  and Hao, Hongxia and Li, Jielan and Yang, Chunlei and
                  Li, Wenjie and Tomioka, Ryota and Xie, Tian},
  title        = {A generative model for inorganic materials design},
  journal      = {Nature},
  volume       = {639},
  pages        = {624--632},
  year         = {2025},
  doi          = {10.1038/s41586-025-08628-5},
  url          = {https://doi.org/10.1038/s41586-025-08628-5},
}

@article{pakornchote2024dpcdvae,
  author       = {Pakornchote, Teerachote and Choomphon-anomakhun,
                  Natthaphon and Arrerut, Sorrjit and Atthapak,
                  Chayanon and Khamkaeo, Sakarn and Chotibut, Thiparat
                  and Bovornratanaraks, Thiti},
  title        = {Diffusion probabilistic models enhance variational
                  autoencoder for crystal structure generative modeling},
  journal      = {Scientific Reports},
  volume       = {14},
  pages        = {1275},
  year         = {2024},
  doi          = {10.1038/s41598-024-51400-4},
  url          = {https://doi.org/10.1038/s41598-024-51400-4},
}

@article{hsu2024scoredynamics,
  author       = {Hsu, Tim and Sadigh, Babak and Bulatov, Vasily and
                  Zhou, Fei},
  title        = {Score Dynamics: Scaling Molecular Dynamics with
                  Picosecond Time Steps via Conditional Diffusion Model},
  journal      = {Journal of Chemical Theory and Computation},
  volume       = {20},
  number       = {6},
  pages        = {2335--2348},
  year         = {2024},
  doi          = {10.1021/acs.jctc.3c01361},
  url          = {https://doi.org/10.1021/acs.jctc.3c01361},
}

@article{hsu2024score,
  author       = {Hsu, Tim and Sadigh, Babak and Bertin, Nicolas and
                  Park, Cheol Woo and Chapman, James and Bulatov,
                  Vasily and Zhou, Fei},
  title        = {Score-based denoising for atomic structure
                  identification},
  journal      = {npj Computational Materials},
  volume       = {10},
  number       = {1},
  pages        = {155},
  year         = {2024},
  doi          = {10.1038/s41524-024-01337-z},
  url          = {https://doi.org/10.1038/s41524-024-01337-z},
}

@article{ronne2024surfacediff,
  author       = {R{\o}nne, Nikolaj and Aspuru-Guzik, Al{\'a}n and
                  Hammer, Bj{\o}rk},
  title        = {Generative diffusion model for surface structure
                  discovery},
  journal      = {Physical Review B},
  volume       = {110},
  number       = {23},
  pages        = {235427},
  year         = {2024},
  doi          = {10.1103/PhysRevB.110.235427},
  url          = {https://doi.org/10.1103/PhysRevB.110.235427},
}

@article{kwon2024spectroscopy,
  author       = {Kwon, Hyuna and Hsu, Tim and Sun, Wenyu and Jeong,
                  Wonseok and Aydin, Fikret and Sharma, Sashank and
                  Meng, Fanchen and Carbone, Matthew R. and Chen, Xiao
                  and Lu, Deyu and Cheng, Liwen F. and Heo, Tae Wook
                  and Pham, Tuan Anh},
  title        = {Spectroscopy-guided discovery of three-dimensional
                  structures of disordered materials with diffusion
                  models},
  journal      = {Machine Learning: Science and Technology},
  volume       = {5},
  pages        = {045037},
  year         = {2024},
  doi          = {10.1088/2632-2153/ad8c41},
  url          = {https://doi.org/10.1088/2632-2153/ad8c41},
}

@article{zheng2024amorphousdb,
  author       = {Zheng, Hui and Sivonxay, Eric and Christensen, Rasmus
                  and Gallant, Max and Luo, Ziyao and McDermott,
                  Matthew and Huck, Patrick and Persson, Kristin A.},
  title        = {The \emph{ab initio} non-crystalline structure
                  database: empowering machine learning to decode
                  diffusivity},
  journal      = {npj Computational Materials},
  volume       = {10},
  pages        = {295},
  year         = {2024},
  doi          = {10.1038/s41524-024-01469-2},
  url          = {https://doi.org/10.1038/s41524-024-01469-2},
}

@inproceedings{finkler2025amden,
  title={AMDEN: Amorphous Materials DEnoising Network},
  author={Finkler, Jonas A and Lin, Yan and Smedskjaer, Morten M and Du, Tao and Hu, Jilin},
  booktitle={AI for Accelerated Materials Design-NeurIPS 2025},
  year={2025}
}

@article{yang2025amorphousdiff,
  author       = {Yang, Kai and Schwalbe-Koda, Daniel},
  title        = {A generative diffusion model for amorphous materials},
  journal      = {npj Computational Materials},
  year         = {2025},
  doi          = {10.1038/s41524-025-01901-1},
  url          = {https://doi.org/10.1038/s41524-025-01901-1},
  note         = {In press},
}

@misc{dm2,
  author       = {{Digital Synthesis Lab}},
  title        = {{DM2}: Denoising-Diffusion Model for Amorphous-Oxide Structures},
  year         = {2024},
  howpublished = {\url{https://github.com/digital-synthesis-lab/DM2}},
  note         = {Software; based on the score-based denoising method of Hsu et al.\ (2024)}
}

@misc{li2025amorphouscond,
  author       = {Li, Honglin and Liu, Chuhao and Guo, Yongfeng and
                  Luo, Xiaoshan and Chen, Yijie and Liu, Guangsheng and
                  Li, Yu and Wang, Ruoyu and Wang, Zhenyu and Wu,
                  Jianzhuo and Ma, Cheng and Xie, Zhuohang and Lv,
                  Jian and Ding, Yufei and Zhang, Huabin and Luo, Jian
                  and Zhong, Zhicheng and Li, Mufan and Wang, Yanchao
                  and Li, Wan-Lu},
  title        = {Conditional Generative Modeling for Amorphous
                  Multi-Element Materials},
  year         = {2025},
  eprint       = {2503.07043},
  archivePrefix= {arXiv},
  primaryClass = {cond-mat.mtrl-sci},
  url          = {https://arxiv.org/abs/2503.07043},
}

@article{mavracic2018tio2,
  author       = {Mavra{\v c}i{\'c}, Juraj and Mocanu, Felix C. and
                  Deringer, Volker L. and Cs{\'a}nyi, G{\'a}bor and
                  Elliott, Stephen R.},
  title        = {Similarity Between Amorphous and Crystalline Phases:
                  The Case of {TiO\textsubscript{2}}},
  journal      = {Journal of Physical Chemistry Letters},
  volume       = {9},
  number       = {11},
  pages        = {2985--2990},
  year         = {2018},
  doi          = {10.1021/acs.jpclett.8b01067},
  url          = {https://doi.org/10.1021/acs.jpclett.8b01067},
}

@article{harrelson2021niobium,
  author       = {Harrelson, Thomas F. and Sheridan, Evan and Kennedy,
                  Ellis and Vinson, John and N'Diaye, Alpha T. and
                  Alto{\'e}, M. Virginia P. and Schwartzberg, Adam and
                  Siddiqi, Irfan and Ogletree, D. Frank and Scott,
                  Mary C. and Griffin, Sin{\'e}ad M.},
  title        = {Elucidating the local atomic and electronic structure
                  of amorphous oxidized superconducting niobium films},
  journal      = {Applied Physics Letters},
  volume       = {119},
  number       = {24},
  pages        = {244004},
  year         = {2021},
  doi          = {10.1063/5.0069428},
  url          = {https://doi.org/10.1063/5.0069428},
}

@inproceedings{peebles2023dit,
  author       = {Peebles, William and Xie, Saining},
  title        = {Scalable Diffusion Models with Transformers},
  booktitle    = {Proceedings of the IEEE/CVF International Conference
                  on Computer Vision (ICCV)},
  pages        = {4195--4205},
  year         = {2023},
  eprint       = {2212.09748},
  archivePrefix= {arXiv},
  url          = {https://arxiv.org/abs/2212.09748},
}

@inproceedings{esser2024sd3,
  author       = {Esser, Patrick and Kulal, Sumith and Blattmann,
                  Andreas and Entezari, Rahim and M{\"u}ller, Jonas and
                  Saini, Harry and Levi, Yam and Lorenz, Dominik and
                  Sauer, Axel and Boesel, Frederic and Podell, Dustin
                  and Dockhorn, Tim and English, Zion and Lacey, Kyle
                  and Goodwin, Alex and Marek, Yannik and Rombach,
                  Robin},
  title        = {Scaling Rectified Flow Transformers for
                  High-Resolution Image Synthesis},
  booktitle    = {International Conference on Machine Learning (ICML)},
  year         = {2024},
  eprint       = {2403.03206},
  archivePrefix= {arXiv},
  url          = {https://arxiv.org/abs/2403.03206},
}

@article{choudhary2023alignnff,
  author       = {Choudhary, Kamal and DeCost, Brian and Major, Lily and
                  Butler, Keith and Thiyagalingam, Jeyan and Tavazza,
                  Francesca},
  title        = {Unified graph neural network force-field for the
                  periodic table: solid state applications},
  journal      = {Digital Discovery},
  volume       = {2},
  number       = {2},
  pages        = {346--355},
  year         = {2023},
  doi          = {10.1039/D2DD00096B},
  url          = {https://doi.org/10.1039/D2DD00096B},
}

@misc{graphite,
  author       = {Hsu, Tim and {Lawrence Livermore National Laboratory}},
  title        = {graphite: a graph neural network library for atomistic
                  modeling},
  year         = {2024},
  howpublished = {\url{https://github.com/llnl/graphite}},
  note         = {LLNL-CODE-836648},
}

@article{schwalbekoda2025information,
  author       = {Schwalbe-Koda, Daniel and Hamel, Sebastien and
                  Sadigh, Babak and Zhou, Fei and Lordi, Vincenzo},
  title        = {Model-free estimation of completeness, uncertainties,
                  and outliers in atomistic machine learning using
                  information theory},
  journal      = {Nature Communications},
  volume       = {16},
  pages        = {3962},
  year         = {2025},
  doi          = {10.1038/s41467-025-59232-0},
  url          = {https://doi.org/10.1038/s41467-025-59232-0},
}

@inproceedings{theis2016note,
  author       = {Theis, Lucas and van den Oord, A{\"a}ron and Bethge, Matthias},
  title        = {A note on the evaluation of generative models},
  booktitle    = {International Conference on Learning Representations (ICLR)},
  year         = {2016},
  eprint       = {1511.01844},
  archivePrefix = {arXiv},
  primaryClass = {stat.ML},
  url          = {https://arxiv.org/abs/1511.01844}
}

@misc{kaplan2020scaling,
  author       = {Kaplan, Jared and McCandlish, Sam and Henighan, Tom and
                  Brown, Tom B. and Chess, Benjamin and Child, Rewon and
                  Gray, Scott and Radford, Alec and Wu, Jeffrey and
                  Amodei, Dario},
  title        = {Scaling Laws for Neural Language Models},
  year         = {2020},
  eprint       = {2001.08361},
  archivePrefix = {arXiv},
  primaryClass = {cs.LG},
  url          = {https://arxiv.org/abs/2001.08361}
}

@inproceedings{hoffmann2022training,
  author       = {Hoffmann, Jordan and Borgeaud, Sebastian and Mensch, Arthur
                  and Buchatskaya, Elena and Cai, Trevor and Rutherford, Eliza
                  and de Las Casas, Diego and Hendricks, Lisa Anne and
                  Welbl, Johannes and Clark, Aidan and Hennigan, Tom and
                  Noland, Eric and Millican, Katie and van den Driessche, George
                  and Damoc, Bogdan and Guy, Aurelia and Osindero, Simon and
                  Simonyan, Karen and Elsen, Erich and Rae, Jack W. and
                  Vinyals, Oriol and Sifre, Laurent},
  title        = {Training Compute-Optimal Large Language Models},
  booktitle    = {Advances in Neural Information Processing Systems (NeurIPS)},
  year         = {2022},
  eprint       = {2203.15556},
  archivePrefix = {arXiv},
  primaryClass = {cs.CL},
  url          = {https://arxiv.org/abs/2203.15556}
}

@misc{ho2022classifierfree,
  author       = {Ho, Jonathan and Salimans, Tim},
  title        = {Classifier-Free Diffusion Guidance},
  year         = {2022},
  eprint       = {2207.12598},
  archivePrefix = {arXiv},
  primaryClass = {cs.LG},
  url          = {https://arxiv.org/abs/2207.12598}
}

@article{stillinger1995topographic,
  author       = {Stillinger, Frank H.},
  title        = {A Topographic View of Supercooled Liquids and Glass Formation},
  journal      = {Science},
  volume       = {267},
  number       = {5206},
  pages        = {1935--1939},
  year         = {1995},
}

@article{debenedetti2001supercooled,
  author       = {Debenedetti, Pablo G. and Stillinger, Frank H.},
  title        = {Supercooled liquids and the glass transition},
  journal      = {Nature},
  volume       = {410},
  pages        = {259--267},
  year         = {2001},
}

@article{vajente2021tigeo2,
  author       = {Vajente, G. and Yang, L. and Davenport, A. and Fazio, M. and
                  Ananyeva, A. and Zhang, L. and Billingsley, G. and
                  Prasai, K. and Markosyan, A. and Bassiri, R. and
                  Fejer, M. M. and Chicoine, M. and Schiettekatte, F. and
                  Menoni, C. S.},
  title        = {Low Mechanical Loss \ce{TiO2}:\ce{GeO2} Coatings for Reduced
                  Thermal Noise in Gravitational Wave Interferometers},
  journal      = {Physical Review Letters},
  volume       = {127},
  pages        = {071101},
  year         = {2021},
  doi          = {10.1103/PhysRevLett.127.071101},
}

@article{harry2007titania,
  author       = {Harry, Gregory M. and Abernathy, Matthew R. and
                  Becerra-Toledo, Andres E. and Armandula, Helena and
                  Black, Eric and Dooley, Kate and Eichenfield, Matt and
                  Nwabugwu, Chinyere and Villar, Akira and
                  Crooks, D. R. M. and Cagnoli, Gianpietro and
                  Hough, Jim and How, Charles R. and MacLaren, Ian and
                  Murray, Peter and Reid, Stuart and Rowan, Sheila and
                  Sneddon, Peter H. and Fejer, Martin M. and Route, Roger and
                  Penn, Steven D. and Ganau, Pierre and Mackowski, Jean-Marie
                  and Michel, Christophe and Pinard, Laurent and
                  Remillieux, Alban},
  title        = {Titania-doped tantala/silica coatings for gravitational-wave
                  detection},
  journal      = {Classical and Quantum Gravity},
  volume       = {24},
  number       = {2},
  pages        = {405--415},
  year         = {2007},
}

@incollection{suito1992geo2,
  author    = {Suito, K. and Miyoshi, M. and Sasakura, T. and Fujisawa, H.},
  title     = {Elastic Properties of Obsidian, Vitreous {SiO$_2$}, and Vitreous
               {GeO$_2$} under High Pressure up to 6 {GPa}},
  booktitle = {High-Pressure Research: Application to Earth and Planetary Sciences},
  series    = {Geophysical Monograph Series},
  volume    = {67},
  pages     = {219--225},
  publisher = {American Geophysical Union},
  address   = {Washington, D.C.},
  year      = {1992},
  doi       = {10.1029/GM067p0219}
}

@article{anderson1997tio2,
  author  = {Anderson, O. and Ottermann, C. R. and Kuschnereit, R. and
             Hess, P. and Bange, K.},
  title   = {Density and {Young}'s modulus of thin {TiO$_2$} films},
  journal = {Fresenius' Journal of Analytical Chemistry},
  volume  = {358},
  pages   = {315--318},
  year    = {1997},
  doi     = {10.1007/s002160050416}
}

@article{rouxel2007glass,
  author  = {Rouxel, Tanguy},
  title   = {Elastic Properties and Short-to Medium-Range Order in Glasses},
  journal = {Journal of the American Ceramic Society},
  volume  = {90},
  number  = {10},
  pages   = {3019--3039},
  year    = {2007},
  doi     = {10.1111/j.1551-2916.2007.01945.x}
}

\end{document}